\documentclass[fleqn,usenatbib]{mnras}

\usepackage{newtxtext,newtxmath}

\usepackage[T1]{fontenc}

\DeclareRobustCommand{\VAN}[3]{#2}
\let\VANthebibliography\thebibliography
\def\thebibliography{\DeclareRobustCommand{\VAN}[3]{##3}\VANthebibliography}

\usepackage{graphicx}	
\usepackage{amsmath}	
\usepackage{newtxtext,newtxmath}

\newcommand{\LL}{\mathcal{L}}

\newcommand{\nenya}{{\tt Nenya }}
\newcommand{\vilya}{{\tt Vilya }}
\newcommand{\narya}{{\tt Narya }}
\newcommand{\theone}{{\tt TheOne }}

\newcommand{\Msun}{ h^{-1}{\rm M_{ \odot}}}
\newcommand{\hMpc}{ h^{-1}{\rm Mpc}}

\newcommand{\ihMpc}{ h\,{\rm Mpc}^{-1}}

\usepackage[dvipsnames]{xcolor}
\definecolor{mygreen}{rgb}{0.1, 0.75, 0.3}
\definecolor{myorange}{rgb}{0.7, 0.2, 0.}

\title[Emulating RSD biased tracers]{The Bacco Simulation Project: Bacco Hybrid Lagrangian Bias Expansion Model in Redshift Space}

\author[Pellejero Iba\~nez et al.]{
Marcos Pellejero Iba\~nez,$^{1}$\thanks{E-mail: mpellejero@dipc.org}
Raul E. Angulo,$^{1,2}$ 
Matteo Zennaro,$^{1}$
Jens St\"ucker,$^{1}$
Sergio Contreras,$^{1}$ \newauthor
Giovanni Aric\`o,$^{3}$
and Francisco Maion$^{1}$
\\
$^{1}$Donostia International Physics Center (DIPC), Paseo Manuel de Lardizabal, 4, 20018 Donostia-San Sebasti\'an, Spain.\\
$^{2}$IKERBASQUE, Basque Foundation for Science, 48013, Bilbao, Spain.\\
$^{3}$Institute for Computational Science, University of Zurich, Winterthurerstrasse 190, 8057 Zurich, Switzerland.
}

\date{Accepted XXX. Received YYY; in original form ZZZ}

\pubyear{2022}

\begin{document}
\label{firstpage}
\pagerange{\pageref{firstpage}--\pageref{lastpage}}
\maketitle

\begin{abstract}
We present an emulator that accurately predicts the power spectrum of galaxies in redshift space as a function of cosmological parameters. Our emulator is based on a 2nd-order Lagrangian bias expansion that is displaced to Eulerian space using cosmological $N$-body simulations. Redshift space distortions are then imprinted using the  non-linear velocity field of simulated particles and haloes. We build the emulator using a forward neural network trained with the simulations of the BACCO project, which covers an 8-dimensional parameter space including massive neutrinos and dynamical dark energy. We show that our emulator provides unbiased cosmological constraints from the monopole, quadrupole, and hexadecapole of a mock galaxy catalogue that mimics the BOSS-CMASS sample down to nonlinear scales ($k\sim0.6\ihMpc$). This work opens up the possibility of robustly extracting cosmological information from small scales using observations of the large-scale structure of the Universe. 
\end{abstract}




\section{Introduction}

One of the main probes of the physics of the Universe is the measurement of the growth rate of structure via redshift space distortions (RSD). RSD refer to the apparent anisotropy in the clustering of galaxies caused by their peculiar velocity.  \citep[e.g.][]{1987MNRAS.227....1K,2013PhRvD..88f3537D,2013MNRAS.432.1928T,2015MNRAS.454.4326P}. Growth rate measurements could tighten constraints on cosmological parameters, detect the signature of massive neutrinos, and even distinguish among competing gravity theories. For this reason, RSD measurements will be carried at high precision by future large-scale-structure (LSS) surveys (e.g. Euclid, \citealt{Laurejis2011} and \citealt{Euclid}, and DESI, \citealt[][]{DESI}). However, to fully realise the potential of RSD for cosmology, it is necessary to develop highly-accurate theoretical models to interpret future data.

The development of RSD models is an active field of research where multiple analytical approachess have been proposed (e.g. \citealt{Fisher1995}, \citealt{Hamana2003}, \citealt{Scoccimaro2004}, \citealt{TaruyaAtsushiNishimichi2010}, \citealt{Seljak_2011}, \citealt{JenningsBaughPascoli2011}, \citealt{ReidWhite2011}, \citealt{BianchiPercivalBel2016}, \citealt{VlahCastorinaWhite2016}, \citealt{KuruvillaPorciani2018}, \citealt{Cuesta2020}, \citealt{Chen2020}). Unfortunately, none of these methods can accurately describe clustering measurements on intermediate and small scales. For instance, models typically fail on scales smaller than  $k\approx 0.2-0.3h/$Mpc: BOSS (see e.g. \citealt{BOSS2017}, \citealt{Chuang_2017}, \citealt{Pellejero-Ibanez_2017} and \citealt{Beutler2017} ), eBOSS (see .g. \citealt{eBOSS2020}), VIPERS (\citealt{VIPERS2017}), GAMA (\citealt{GAMA2013}), the 6 degree Field Galaxy Survey 6dFGS (\citealt{6dF2012}), the Subaru FMOS galaxy redshift survey (\citealt{Subaru2016}) and the WiggleZ Dark Energy Survey (\citealt{WiggleZ2011}). This means that small scales are neglected in the cosmological analyses of state-of-the-art surveys with the corresponding loss of information \citep[see  e.g][]{Quijote2020}.

The main difficulty in building accurate RSD models is the nonlinearity of the physics involved. Gravitational evolution quickly imprints strong nonlinear features in the density and velocity fields, which can only be described by analytic models on large scales where perturbation theory is valid. In addition, describing the mapping between matter and observed LSS tracers remains a challenge -- this relation is expected to depend on the details of galaxy formation physics, which is extremely complex and still has several uncertain aspects.

A possible approach is to build theoretical models using  numerical simulations which can accurately follow these nonlinear processes \citep[see][for a review]{AnguloHahn2022}. For instance, current hydro-dynamical simulations produce realistic populations of galaxies and clusters (see e.g. \citealt{Vogelsberger2013}, \citealt{Schaye2015}, \citealt{McCarthy2017}), and computational power has allowed them to be carried over multiple combinations of cosmological and galaxy formation parameters albeit over small cosmic volumes (e.g. \citealt{Camels2022}). Alternative ways of describing the galaxy-matter connection are (semi)empirical methods built on top of gravity-only simulations. For instance, halo occupation distribution-based approaches (HOD) assume that the mass of a dark matter halo fully determines the number of central and satellite galaxies it contains. Other approaches are based on the SubHalo Abundance Matching Technique which employ physical properties of dark matter subhalos to assign them physical properties of a galaxy population, such as stellar mass or formation rate. In fact, there has been multiple works placing cosmological constraints using clustering small scales (see e.g. \citealt{Reid2014}, \citealt{Chapman2021} \citealt{Lange2022}, \citealt{Zhai2022}, \citealt{Kobayashi2022}, \citealt{Yuan2022})

The main shortcoming of such approaches -- hydrodynamic simulations, as well as HODs and SHAMs -- is that it is necessary to make (not always easily testable) assumptions regarding galaxy-halo connections and galaxy evolution physics. For instance: the specific implementations of black hole growth and feedback, the functional form of the HOD, or, more generally, the completeness and accuracy of the model. In fact, it has been shown that cosmological constraints can be severely biased depending on the choice of sub-grid physics or hydrodynamical code (\citealt{Villaescusa-Navarro2021}). 

A more agnostic approach can be taken. The galaxy-matter connection can be parametrised by a series of functional dependencies -- the bias expansion -- informed by the symmetries of the underlying laws rather than the details of the galaxy formation physics \citep{Desjacques2018}. This approach has been traditionally adopted in combination with pertubative descriptions of structure formation (see e.g. \citealt{BernardeauEtal2002, VlahCastorinaWhite2016, Ivanov_2020, d_Amico_2020, Colas_2020, ChudaykinEtal2020, Kitaura2020}). 

Recently, a pertubative bias expansion has been built on top of matter statistics computed directly from $N$-body simulation \citep{Modi_2020,ZennaroAnguloPellejero2021,Kokron2021}, which seems to extend its range of applicability. Specifically, in \cite{ZennaroAnguloPellejero2021} we constructed an emulator for the real-space power spectrum of biased tracers. In addition, in \cite{ZennaroAnguloContreras2021} we showed that the model recovered accurate results of the two-point statistics of thousands of mock galaxy samples down to scales $k\approx 0.7 h/$Mpc. This accuracy held for stellar-mass and star formation rate selected galaxy samples at various number densities and redshifts. This so-called Hybrid approach was recently employed to model 3x2 point measurements of the DES survey \citep{hefty}. 

In \cite{PellejeroIbanez2021} we expanded these ideas and showed that an analogous model can be build to model RSD by also including the velocity field from $N$-body simulations.
At a fixed cosmology, this model described, with sub-percent accuracy, the redshift space power spectrum down to scales of $k\approx 0.6 h/$Mpc -- approximately extending by a factor of 2 the smallest scale included in earlier perturbative RSD models. Perhaps more importantly, the recovered bias parameters were in agreement with those estimated in real space, which suggested that our RSD modelling is complete and consistent with the real space one. In this paper, we build an emulator for the model described in \cite{PellejeroIbanez2021} to incorporate the cosmological parameter dependencies. We will then show that this model can indeed extract unbiased cosmological constraints from realistic mock galaxy samples, which paves the way for its use in real data analyses.

We divide this paper as follows. In \S\ref{sec:model} we discuss our model for the redshift-space clustering of biased tracers based on $N$-body simulations. In \S\ref{sec:emulation} we present and discuss the specifics of our emulator technique, the re-scaling algorithm and the feed forward neural network, introducing the main $N$-body simulations used to build the theoretical model. In \S\ref{sec:validation} we validate the model by testing its cosmology dependence and assessing its error. In \S\ref{sec:application} we test the emulator on a controlled case given by a BOSS-CMASS galaxy mock and show the recovered cosmological parameters. We conclude and summarise our results in \S\ref{sec:conclusion}. 

\section{Hybrid $N$-body bias RSD model}
\label{sec:model}

In this section, we present our model for the two-point function of galaxies (or tracers in general) in redshift space. We closely follow the methodology of \cite{PellejeroIbanez2021} to which we refer the reader for additional information. 

\subsection{Real space}

The modelling of the clustering statistics in the so called ``hybrid'' approaches \citep{Modi_2020} contains two ingredients: i) a map from Lagrangian, $\pmb{q}$,  to Eulerian, $\pmb{x}$, space using the displacement field as measured on $N$-body simulations, $\pmb{\psi}(\pmb{q})$, and ii) a functional relation between the density field of matter and that of galaxies i.e. a bias model, $F(\delta_{\rm L}(\pmb{q}))$. 

Regarding the first ingredient, we can write the Lagrangian to Eulerian space mapping as

\begin{equation}
\begin{split}
    \pmb{x}=\pmb{q}+\pmb{\psi}(\pmb{q})\, ,
\end{split}
\end{equation}

\noindent which defines completely the Eulerian matter overdensity field:

\begin{equation}
\begin{split}
    1+\delta(\pmb{x})=\int \rm{d}^3q \, \delta_{\rm D}(\pmb{x}-\pmb{q}-\pmb{\psi}(\pmb{q}))\, .
    \label{eq:DMmapp}
\end{split}
\end{equation}

\noindent This equation states that the mass overdensity is simply the result of the deformation of Lagrangian volume elements. $\pmb{\psi}(\pmb{q})$ can be easily measured in $N$-body simulations by comparing the Lagrangian position of simulation particles and their position in a snapshot at any desired redshift.

Regarding the second ingredient, we employ a 2nd-order Lagrangian bias model \citep{PhysRevD.78.083519,Desjacques2018}:

\begin{equation}
\begin{split}
    w(\pmb{q}) = F(\delta_{\rm L}(\pmb{q})) = & 1 + b_1\delta_{\rm{L}}(\pmb{q}) + b_2 \left( \delta^2_{\rm{L}}(\pmb{q})-\langle\delta^2_{\rm{L}}(\pmb{q})\rangle \right) \\ & + b_s \left( s^2(\pmb{q}) - \langle s^2(\pmb{q})\rangle \right) + b_{\nabla} \nabla^2 \delta_{\rm{L}}(\pmb{q}) \; .
	\label{eq:model}
\end{split}
\end{equation}

\noindent Here $\delta_{\rm{L}}(\pmb{q})$ stands for the linear field and $s^2$ is the traceless part of the tidal field, $s^2=s_{ij}s^{ij}=(\partial_i\partial_j\phi(\pmb{q}) - 1/3\delta^{\rm{K}}_{ij}\delta_{\rm{L}}(\pmb{q}))^2$, with $\phi(\pmb{q})$ the linear gravitational potential, $\nabla^2\phi = 4\pi G \bar{\rho}\delta_{\rm{L}}(\pmb{q})$. Following \cite{Modi_2020} we can now write, analogously to Eq.~\ref{eq:DMmapp},

\begin{equation}
\begin{split}
    1+\delta_{\rm tr}(\pmb{x})=\int \rm{d}^3q \, w(\pmb{q}) \, \delta_{\rm D}(\pmb{x}-\pmb{q}-\pmb{\psi}(\pmb{q}))\, ,
    \label{eq:galmapp}
\end{split}
\end{equation}

\noindent where the subscript ``tr'' applies to a generic tracer of the density field. The function $w(\pmb{q})$ weights the importance of different Lagrangian fields in representing the density of tracers at a given $\pmb{x}$. 

As stated previously, these hybrid models have been tested multiple times \citep{Modi_2020,Kokron2022}. An important step forward was constructing an emulator which allowed sampling cosmological parameters \citep{ZennaroAnguloPellejero2021,Kokron2021}. Additionally, in \cite{ZennaroAnguloContreras2021} we explored the validity of the model against thousands of different galaxy catalogues built with different values for the free parameter of the underlying galaxy model, and then quantified the relationship between between bias parameters.

\subsection{Redshift space}

To model clustering in redshift space within hybrid models, we must account for the anisotropy caused by the peculiar velocity of biased tracers, $\pmb{v}$. Specifically, the observed line-of-sight position, $\pmb{s}$, is related to the real space position, $\pmb{x}$, via:

\begin{equation}
    \pmb{s}=\pmb{x}+\frac{\hat{\pmb{z}} \cdot \pmb{v}(\pmb{x})}{aH} \, \hat{\pmb{z}} \; .
	\label{eq:RSD}
\end{equation}

\noindent where $\hat{\pmb{z}}$ is the line-of-sigh direction, $a$ stands for the scale factor, and $H$ for the Hubble function. Following \cite{Pellejero-Ibanez2020}, we include this effect by breaking the symmetry of the displacement field, without modifying the bias expansion. Eq.~\ref{eq:galmapp} becomes

\begin{equation}
    1+\delta^s_{\rm tr}(\pmb{s})=\int \rm{d}^3q \, w(\pmb{q}) \, \delta_{\rm D}(\pmb{s}-\pmb{q}-\pmb{\psi}^s(\pmb{q}))\, ,
    \label{eq:galmappRSD}
\end{equation}

\noindent with $\pmb{\psi}^s(\pmb{q})$ accounting for the shifts in the line-of-sight direction due to the velocity field as follows,

\begin{equation}
\begin{split}
    \pmb{\psi}^s(\pmb{q}_z) = \pmb{\psi}(\pmb{q}_z) + \frac{\hat{\pmb{q}}_z \cdot \pmb{v}_{\rm{tr}}(\pmb{q})}{aH} \, \hat{\pmb{q}}_z \, ,
    \label{eq:galDispRSD}
\end{split}
\end{equation}

\noindent where we define $\pmb{v}_{\rm{tr}}$ based on the velocities provided by the simulation. 

Since the real-space displacement is determined by dark matter particles in an $N$-body simulation, it would seem natural to set $\pmb{v}_{\rm{tr}}$ as the velocity from the same particles. However, this would be incorrect. Compared to galaxies, dark matter exhibits different velocity dispersion profiles in halos \citep[e.g.][]{Wu2013}. This is result of galaxies sampling a coarsed-grained version of the underlying phase-space distribution function. The details of this average depends on the details of the sample and on galaxy formation physics.

In this paper, we employ a model that can adapt to the velocity statistics of various sets of tracers. Our first step is to define $\pmb{v}_{\rm{tr}}(\pmb{x})$ in terms of the halo and particle velocities of the simulation

\begin{equation}
\pmb{v}_{\rm{tr}}(\pmb{x}) = \begin{cases} \pmb{v}(\pmb{x}) , & \mbox{if the tracer is outside of a halo} \\ \pmb{v}_{\rm{halo}}(\pmb{x}), & \mbox{if the tracer is inside of a halo} \end{cases}
\label{eq:velocity}
\end{equation}

\noindent where $\pmb{v}$ is the velocity vector as sampled by simulation particles, and $\pmb{v}_{\rm{halo}}$ is the center of mass velocity of the halo containing the particle at location $\pmb{x}$.
We expect this to work well for the redshift space distortions of central galaxies since they tend to be at rest relative to their parent halo. \footnote{Note this is not strictly true as central galaxies could display a velocity relative to the halo as large as $\sim30$\% of the halo velocity dispersion \citep{Guo2015}. However, this value depends on the exact definition of halo velocity.}

This distortion roughly accounts for the so called ``Kaiser effect'' \citep{1987MNRAS.227....1K} but also incorporates additional contributions resulting from the non-linearity of the halo velocity. To include small-scale non-linearities, manifested as ``Fingers of God'' (FoG), a standard approach is to use a streaming model (see e.g. \citealt{ReidWhite2011}  and \citealt{VlahCastorinaWhite2016}) in configuration space, where linear theory is spliced together with an approximation for random motion of particles in collapsed objects. We include the FoG effect via a convolution with an exponential function \citep{1983ApJ...267..465D}. Since galaxies populate halos of a wide range of masses and velocity dispersions \citep{White2001}, an exponential function seems to work well for describing number density cuts. This means that the velocity distribution of galaxies relative to central halos can be well described by

\begin{align}
  p(v_z) &= (1 - f_{\rm{sat}}) \delta_{\rm{D}} (v_z) + f_{\rm{sat}} \exp \left( -\lambda v_z \right) \; ,
\end{align}

\noindent where $\delta_{\rm{D}}$ is the Dirac-delta distribution accounting for the fact that a fraction of the galaxies ($1 - f_{\rm{sat}}$) are centrals assumed to have zero velocity with respect to haloes. Here, $f_{\rm{sat}}$ refers to the satellite fraction and $\lambda_{\rm{FoG}}$ models the amount of damping due to peculiar velocities, as discussed in \cite{10.1093/mnras/stx3349}. Note that the larger $\lambda_{\rm{FoG}}$, the weaker the FoG effect. Hence, we can model the effect of the intra-cluster velocities onto the redshift-space galaxy field by applying a convolution along the line-of-sight direction:

\begin{equation}
\delta^{s}_{\rm{tr}}(\pmb{s}) \rightarrow \delta^s_{\rm{tr}}(\pmb{s}) \boldsymbol{\ast}_z \left[ (1-f_{\rm{sat}}) \delta_{\rm{D}}(s_z) +f_{\rm{sat}}\exp \left( {-\lambda_{\rm{FoG}} s_z} \right) \right] \; .
\label{eq:densFoG}
\end{equation}
We use the notation $\boldsymbol{\ast}_z$ for the 1D convolution operation in the $z$-direction.

\subsection{Power spectrum in redshift space}

The power spectrum of the tracer field, $\delta^s_{\rm{tr}}$, is defined as

\begin{align}
    \langle \delta_{\rm tr}(k,\mu)\delta_{\rm tr}(k^{\prime},\mu^{\prime}) \rangle = \frac{1}{(2\pi)^3} \delta_{D}(k-k^{\prime})\delta_{D}(\mu-\mu^{\prime})P_{\rm tr}(k,\mu) \; ,
\label{eq:2D_pk}
\end{align}

\noindent where $k\equiv |\pmb{k}|$ and $\mu \equiv \pmb{k} \cdot \pmb{k_z}$. This two-dimensional power spectrum, $P^{\rm FoG}_{\rm tr}$, can be computed as the weighted sum of 15 terms appearing in the bias expansion of Eq.~\ref{eq:model}:
\begin{equation}
\begin{split}
&P_{\rm tr}(k,\mu) = \left( (1-f_{\rm{sat}})+ f_{\rm{sat}} \frac{\lambda^2_{\rm{FoG}}}{\lambda^2_{\rm{FoG}}+k^2\mu^2} \right)^2 \times \\ 
& \Big[ P^s_{11}+2 b_{1} P^s_{1 \delta}+2 b_{2} P^s_{1 \delta^{2}}+2 b_{s^{2}} P^s_{1 s^{2}}+2 b_{\nabla^{2} \delta} P^s_{1 \nabla^{2} \delta}+ \\
&\left(b_{1}\right)^{2} P^s_{\delta \delta}+2 b_{1} b_{2} P^s_{\delta \delta^{2}}+2 b_{1} b_{s^{2}} P^s_{\delta s^{2}}+2 b_{1} b_{\nabla^{2} \delta} P^s_{\delta \nabla^{2} \delta}+ \\
&\left(b_{2}\right)^{2} P^s_{\delta^{2} \delta^{2}}+2 b_{2} b_{s^{2}} P^s_{\delta^{2} s^{2}}+2 b_{2} b_{\nabla^{2} \delta} P^s_{\delta^{2} \nabla^{2} \delta}+ \\
&\left(b_{s^{2}}\right)^{2} P^s_{s^{2} s^{2}}+2 b_{s^{2}} b_{\nabla^{2} \delta} P^s_{s^{2} \nabla^{2} \delta}+\left(b_{\nabla^{2} \delta}\right)^{2} P^s_{\nabla^{2} \delta \nabla^{2} \delta}+ \rm{Noise} \Big] \, .
\end{split}
\label{eq:Pkmodel}
\end{equation}

\noindent Here, the $P^s_{ij}(k,\mu)$ (we dropped the $k$ and $\mu$ dependence for brevity) with $i \in \{1, \delta, \delta^2, s^2\, \nabla\delta \}$ are the redshift-space auto and cross power spectra of the bias expansion measured from the $N$-body simulation. Each of these spectra is computed using redshift-space displacement fields according to Eqs.~\ref{eq:galmappRSD}-\ref{eq:velocity}. These spectra are the quantities that we will later emulate. 

In previous works, the stochastic contribution to the power spectrum was modelled as $A_{\rm noise}/\bar{n}$,  where $A_{\rm noise}$ is a free parameter and $\bar{n}$ is the mean number density of the sample (see e.g. \citealt{ZennaroAnguloPellejero2021}). However, the stochastic component could be in fact much more complex as it might contain the contribution of terms not included in our bias expansion and of unmodelled small-scale physics. Indeed a constant term corresponds to the lowest-order expansion of the stochastic term, and the next non-vanishing order scales as a $k^2$ \citep{Perko2016}. For this reason, we will also consider this additional term:

\begin{equation}
{\rm Noise} = \frac{1}{\bar{n}_{\rm tr}}\left( \epsilon_1+\epsilon_2k^2 \right) ,
\label{eq:Noise}
\end{equation}

\noindent where $\epsilon_1$ and $\epsilon_2$ and two free parameters. Note that, in redshift space, this term does not have line-of-sight dependence. Thus, it affects only the monopole of $P_{\rm tr}$. In principle, this stochastic contribution is different in velocity space, which appears at lowest order as a  $k^2\mu^2$ term. Here, we have not included this term as it does not seem to be required by the particular galaxy samples we study.

In summary, our model operates as follows: i) advecting tracer particles to Eulerian space by using the non-linear redshift-space displacement field from the $N$-body simulation, ii) constructing a galaxy density field by weighting each volume element in Lagrangian space with a function of the initial linear fields through a small set of free bias parameters, and iii) modelling intra-halo velocities (FoG) with two additional parameters.

Hence, the total number of free parameters will sum up to 8, $\{b_1,b_2,b_{s^2},b_{\nabla^2\delta}, \lambda_{\rm FoG}, f_{\rm sat},\epsilon_1, \epsilon_2\}$. These can be thought as nuisance parameters -- as in previous data analysis  \citep[e.g.][]{Pellejero-Ibanez_2017} -- or can be studied to understand co-evolution relations -- such as in \citealt{ZennaroAnguloContreras2021}. As shown in \cite{PellejeroIbanez2021}, the described model reproduces the clustering of realistic galaxy mocks with great accuracy for a fixed cosmology.

\section{Building the emulator}
\label{sec:emulation}

As stated in the previous section, our model for the redshift space power spectrum requires 15 cross-spectra of Lagrangian field advected to redshift space using displacement and velocity fields from numerical simulations. The goal of this section is to describe how we make such predictions as a function of cosmology.

Our basic strategy is the following. We employ a suite of high resolution simulations together with cosmology rescaling to densely sample a target cosmological parameter space. Then, we compute Lagrangian fields and the counterparts in redshift space at those cosmologies. Finally, we use this data to train a neural network. As a result, this provides accurate and extremely fast predictions, which makes it possible to use our model in cosmological data analyses.

\begin{table}
  \centering
  \begin{tabular}{cc|cccccccc} 
     \hline
     Cosmology &  $\Omega_{\rm cdm}$ & $\Omega_{\rm b}$ & $h$ & $n_{\rm s}$ & $\sigma_8$\\
     \hline
    \nenya  & 0.265 & 0.050 & 0.60 & 1.01 & 0.9 \\
    \narya  & 0.310 & 0.050 & 0.70 & 1.01 & 0.9 \\
    \vilya  & 0.210 & 0.060 & 0.65 & 0.92 & 0.9 \\
    \theone & 0.259 & 0.048 & 0.68 & 0.96 & 0.9 \\
    \hline
    \theone-BOSS & 0.259 & 0.048 & 0.68 & 0.96 & 0.8 \\
     \hline
     \end{tabular}
    \caption{Cosmological parameters of the four cosmologies simulated in the BACCO project. The parameters not stated here correspond to flat geometry, no massive neutrinos ($M_{\nu}=0$ eV), a dark energy equation of state with $w_0=-1$ and $w_a=0$, and optical depth at recombination $\tau=0.0952$. The lowest row shows the cosmology of the scaled simulation used to produce the galaxy mock.}
  \label{tab:parameters_table}
\end{table}

\subsection{BACCO Simulations}

The BACCO simulations are the core of our biased-tracers emulators. Presented in \cite{AnguloEtal2020}, correspond to 8 large $N$-body simulations designed to cover a wide range of cosmological parameters when combined with cosmology-rescaling. We now provide a summary of their main characteristics.

The BACCO simulations are a set of gravity-only simulations with a box length of $L=1440\,\hMpc$ resolved with $4320^3$ particles, which implies a mass resolution of $m_{\rm p} \sim 3 \times 10^9\,\Msun$. The gravitational evolution was carried out with an updated version of {\tt L-Gadget3} \citep{Springel2005,AnguloEtal2020} used to run the Millennium XXL simulation. These simulations were carried out with 4 distinct sets of cosmological parameters chosen so that they can accurately cover a region of approximately $10\sigma$ around the best fit values obtained by the analysis of the Planck satellite \citep{planck2015-xiii} when combined with cosmology rescaling algorithms \citep{AnguloWhite2010}. The cosmological parameter values are provided in Table \ref{tab:parameters_table}. We refer to \cite{ContrerasAnguloetal2020} for further information on how these parameters were chosen. 

In each of these simulations we employed the ``fixed \& paired'' technique \citep{AnguloPontzen2016}. This methodology requires access to two simulations whose initial Fourier amplitudes have been fixed to the ensemble average, but their initial phases are inverted. This configuration allows for a dramatic reduction of the noise in the halo and matter power spectrum due to the cancellation of cosmic variance -- by more than two orders of magnitude for $k<0.1\ihMpc$  \citep{UNITsim2019, MaionAnguloZennaro2022}.

\subsection{Cosmology scaling}
\label{sec:rescaling}

For any set of cosmological parameters, the model described in \S\ref{sec:model} needs an $N$-body simulation. These simulations are computationally expensive and cannot be used to densely sample a given target cosmological parameter space. Therefore, to capture the cosmology dependence of the power spectra in Eq.~\ref{eq:Pkmodel}, we will employ the cosmology rescaling algorithm of \cite{AnguloWhite2010}. 

This approach exploits the idea that the nonlinear structure in a given cosmology can be mimicked by rescaling the outputs of a simulation carried out in a nearby cosmology. The rescaling is performed by transforming the length unit and by considering a different output time such that the linear variance of the field as a function of scale matches that of the rescaled simulation. Additionally, large-scale modes are modified by displacing a given simulation particle with the difference of the second order Lagrangian Perturbation Theory (2LPT) displacements in the original and target cosmologies. Velocities are modified in an analogous manner. In \cite{ContrerasAnguloetal2020} we showed that the matter power spectrum of dark matter, halos, and subhaloes can be retrieved to better than 3\% up to $k \sim 5\ihMpc$, and to better than 1\% for the real and redshift-space matter power spectrum over the scales we explore in this paper. This approach was employed by \cite{AnguloEtal2020} to construct an emulator for the nonlinear matter power spectrum; by \cite{Zennaro2019} to include the effects of massive neutrinos; by \cite{Arico2020} to incorporate the effects of baryonic physics such as star formation, gas cooling and feedback, on the matter power spectrum; and by \cite{ZennaroAnguloPellejero2021} to construct emulators for the $P_{ij}$ terms in real space.

For a given redshift, this full scaling algorithm takes 20 minutes (on 12 threads using OpenMP parallelisation) to compute the relevant cross-spectra, a negligible amount of time with respect to the typical time demanded by a full $N$-body simulation. Taking as a basis the BACCO simulations presented in Table \ref{tab:parameters_table} we rescale 10 different output times to 400 different cosmological parameters sets distributed according to a Latin-Hypercube \citep{Heitmann2006}. This cosmological space covers 8 parameters of the $w_0w_a$CDM model with massive neutrinos and dynamical dark energy. Specifically, we consider the parameter range:
\begin{eqnarray}
\label{eq:par_range}
\Omega_{\rm m}            &\in& [0.23, 0.4] \nonumber\\
\sigma_8                  &\in& [0.65, 0.9] \nonumber\\
\Omega_b                  &\in& [0.04, 0.06] \nonumber\\
n_s                       &\in& [0.92, 1.01] \nonumber\\
h                         &\in& [0.6, 0.8] \\
M_{\nu} \,[{\rm eV}]      &\in& [0.0, 0.4] \nonumber\\
w_{0}                     &\in& [-1.15, -0.85] \nonumber\\
w_{a}                     &\in& [-0.3, 0.3] \nonumber \\
a                     &\in& [0.2, 1] \nonumber
\end{eqnarray}
where $a$ stands for the expansion factor, $M_\nu$ is the total mass in neutrinos, $\sigma_8$ is the root mean squared linear (dark matter plus baryons) variance in $8\hMpc$ spheres, $\Omega_{\rm m}$ refers to cold matter (cold dark matter plus baryons), and $w_0$ and $w_a$ define the time evolution of the dark energy equation of state through the CPL parametrisation: $w(z) = w_0 + (1 - a) w_a$. We assume a flat geometry, i.e. $\Omega_k = 0$ and $\Omega_m + \Omega_w + \Omega_{\nu} = 1$. We keep fixed the effective number of relativistic species to $N_{\rm eff}=3.046$, and the temperature of the CMB $T_{\rm CMB} = 2.7255$ K and neglect the impact of radiation (i.e. $\Omega_r = 0$).

At each of these 4000 parameter sets, we compute all the 15 $P^s_{ij}(k,\mu)$ terms of Eq.~\ref{eq:Pkmodel}. Even though our methodology increases the speed of the computation of these quantities by several orders of magnitude, it is still too slow for parameter inference: the computation still requires several minutes of CPU time, which would make a traditional MCMC very computationally expensive. As a solution, we propose to use these scaled quantities as training data for a feed-forward neural network emulator. 


\begin{figure*}
\includegraphics[width=\textwidth]{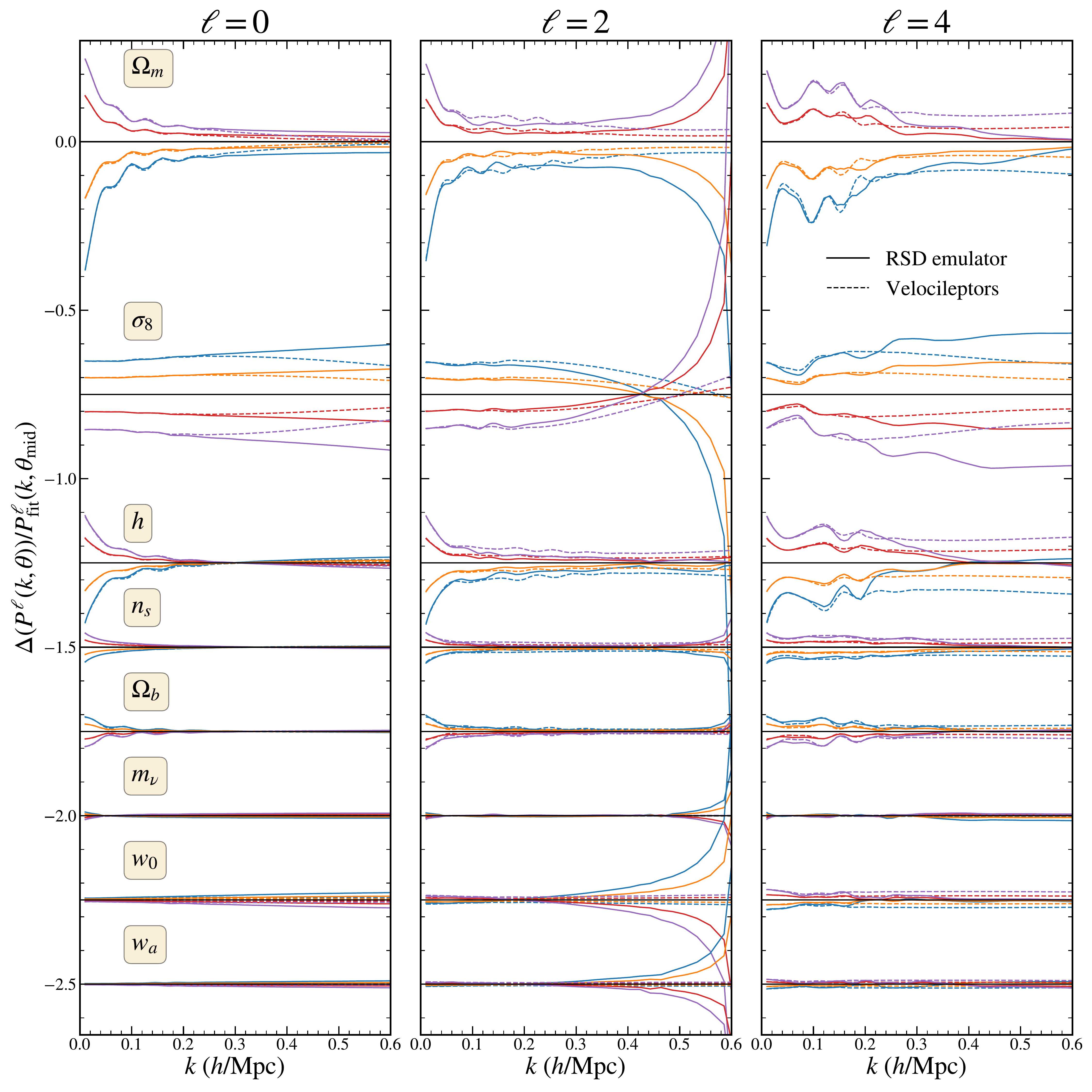}
\caption{Emulator monopole, quadrupole, and hexadecapole dependencies over cosmological parameters for a specified set of bias parameters. Each line represents the fractional difference between the model evaluated at a given cosmology $\theta$ and the model evaluated at the midpoint of the spanned parameter space $\theta_{\rm{mid}}$. The selected bias parameters are those that best accommodate the galaxy mock data at the true cosmology. Only one cosmological parameter is changed at a time. The parameters for this test are $\theta = \theta_{\rm{mid}}\pm N\Delta_{\theta}$, where $\theta_{\rm{mid}}$ is the middle value of the cosmological parameter space described in Eq.~\ref{eq:par_range} and $\Delta_{\theta}=\{0.02, 0.02, 0.002, 0.008, 0.032, 0.04, 0.04, 0.08\}$ are equal steps to cover the prior parameter space starting from the mid-value. We displaced each cosmology by a constant factor to avoid clutter. Dashed lines show the same quantities but for the \texttt{velocileptors} predictions with all counter-terms set to zero.   
\label{fig:emuVSparams}}
\end{figure*}


\subsection{Neural Network emulation}
\label{sec:NNemulation}

Cosmological emulators offer a promising approach to exploit the accuracy of numerical simulation to predict nonlinear cosmological quantities \citep[e.g.][]{Agarwal2014,Kobayashi2020,HeitmannEtal2014,LiuEtal2018,NishimichiEtal2019,DeRoseEtal2019,GiblinEtal2019,EuclidEmu2019,WibkingEtal2019,WintherEtal2019,AnguloEtal2020,EuclidEmu2020, VillaescusaNavarro2021, Jamieson2022}. In such approaches, the results of simulations carried out with various sets of cosmological parameters are interpolated using approximate numerical or analytical functions. A popular technique are feed forward Neural Networks, serving as a highly-precise approximation to almost any continuous function. They are intended to resemble nonlinear functions, such as those describing the summary statistics predicted by simulations. For instance, in \cite{AnguloEtal2020}, we demonstrated that Neural Networks are able to recover the complete nonlinear matter power spectrum interpolated between simulations to an accuracy of a few percent.

Here, we follow the procedure of \cite{AnguloEtal2020}, \cite{Arico2020}, and \cite{ZennaroAnguloPellejero2021}, but applied on the scaled redshift space power spectra $P^s_{ij}(k,\mu)$. However, instead of emulating the 2D power spectra, we emulate their three first non-zero multipoles (monopole, quadrupole and hexadecapole) and then reconstruct $P^s_{ij}(k,\mu)$ as following:

\begin{equation}
    P^{s}_{ij}(k,\mu) \sim P_{ij,0}(k)\mathcal{L}_{0}(\mu) + P_{ij,2}(k)\mathcal{L}_2(\mu) + P_{ij,4}(k)\mathcal{L}_{4}(\mu) \, ,
    \label{eq:multipoles_exp}
\end{equation}

\noindent where $\mathcal{L}_X$ are the Legendre polynomials of order $X$, and the multipoles are defined as

\begin{equation}
P_{ij,\ell}(k) = \frac{2\ell+1}{2}\int_{-1}^{1}d\mu\,P_{ij}^s(k,\mu)\mathcal{L}_\ell(\mu)\, .
\label{eq:pkpoles}
\end{equation}

Naturally, by including higher-order multipoles we could achieve a more accurate reconstruction of $P^{s}(k,\mu)$. However, we found that the contribution of multipoles higher than $\ell=4$ is negligible compared to the cosmic variance of upcoming LSS surveys -- they modify the power spectrum by less than 20\% of expected Gaussian cosmic variance for a 24(Gpc/$h$)$^3$ volume at $k\approx1h/{\rm{Mpc}}$. Nevertheless, it is straightforward to include and emulate higher multipoles in the future if required.

We have found that a feed-forward neural network with a relatively simple architecture is able to produce an accurate emulation of the redshift space multipoles. Specifically, we employ a fully-connected network with 2 hidden layers and 400 neurons each. For the activation function we use the Rectified Linear Units (ReLUs). Using the adaptive stochastic optimization algorithm {\tt Adam} \citep{Kingma2014adam}, the network is trained with a default learning rate of $10^{-3}$. For the implementation of the neural network we made use of the open-source neural-network python library, {\tt tensorflow-Keras} \citep{tensorflow2015-whitepaper}.

We train a separate neural network for each cross-spectrum and multipole, resulting in a total of $15\times3$ distinct networks. To facilitate the training, we reduce the dynamical range spanned by the spectra by emulating the ratio with respect to theoretical expectations:

\begin{equation}
Q_{ij,\ell}(k,z) =\frac{P_{ij,\ell}(k,z)}{P^{\rm theory}_{ij,\ell}(k,z)} .
\label{eq:Q(k)}
\end{equation}

\noindent The term $P^{\rm theory}_{ij,\ell}(k,z)$ corresponds to the predictions by Lagrangian perturbation theory, as implemented by \texttt{velocileptors}\footnote{\url{https://github.com/sfschen/velocileptors}} (\citealt{Chen2020}, \citealt{Chen2021}). Specifically, velocileptors computes each spectrum in one-loop perturbation theory with effective corrections for small scale effects to account for velocity effects by combining velocity statistics and the correlation function in Lagrangian Perturbation Theory in configuration space. In the cases of the matter multipoles ($P_{11,\ell}$ term), instead of using \texttt{velocileptors}, we employ the redshift-space linear theory power spectrum where the Bayonic Acoustic Oscillations (BAO) have been smeared out according to the expectations of perturbation theory (for details, see section 3.1.2 of \citealt{AnguloEtal2020}). This has the advantage of being positive on all scales and that it aids in an accurate emulation of BAO features. 

Despite the large volume of our simulations and of the cosmic variance reduction due to "Fix-and-Paired" sets, the ratios $Q_{ij}$ will still contain noise which could reduce the accuracy of our neural network since it could learn to reproduce spurious features. To reduce this problem, we further decompose each cross-spectrum and multipole using a Principal Component Analysis (PCA), retaining the first six $k$-vectors with the highest eigenvalues. These principal components are then combined to reconstruct the $Q_{ij}$ which are input to the network.

For training and validation, we split our dataset of 15x4000 sets of cross-spectra monopoles, quadrupoles, and hexadecapoles into distinct groups. The training set contains 95\% of the data required to train each of the $15\times3$ emulators on a single Nvidia Quadro RTX 8000 GPU card in approximately 30 minutes. On the same hardware, the evaluation of each emulator takes approximately 0.002 seconds.

A single evaluation of the model, including the call to all 3x15 emulators and the convolution to incorporate the FoG effect, requires 0.3 seconds, making it suitable for traditional inference analysis in a few hours. 

\subsection{Emulation basis \& LPT transition}

As stated in Eq.~\ref{eq:Q(k)}, the dynamical range of the training data is decreased by taking the ratio with respect to perturbation theory predictions. This implies that to recover the complete  model, we must multiply by the same theory prediction evaluated at the target cosmology which can easily dominate the CPU time needed to obtain our model. In \cite{Chen2022} they accelerated the computation of these theory predictions by employing a fast evaluation via Taylor Series, whereas \cite{DeRose2021} proposed using neural network emulators. We adopt this second strategy and create emulators for $P^{\ell_,{\rm theory}}$ employing a similar architecture as the one described in \S\ref{sec:NNemulation} -- a fully-connected network with 2 hidden layers, 35 neurons each, ReLUs as the activation function, and the {\tt Adam} algorithm with a starting learning rate of $10^{-3}$ that was later lowered to $6 \times 10^{-5}$. As training data, we evaluated $P^{\ell_,{\rm theory}}$ on 40000 cosmologies distributed according to a LH over the same cosmological parameter range as our RSD emulator. Since this emulator is based on a larger training sample which is also unaffected by noise, its accuracy is significantly larger than that of the $N$-body, and it contributes insignificantly to the final uncertainty of our model. 

Taking advantage of the fact that the theory predictions perform exceptionally well on the largest scales, we further use these emulators for scales larger than a given $k_{\rm{lim}}$. This helps in reducing the cosmic variance affecting these scales. For the monopole and quadrupole we choose a $k_{\rm{lim}}<0.05h/{\rm{Mpc}}$, while for the hexadecapole, this value is increased to  $k_{\rm{lim}}<0.15h/{\rm{Mpc}}$. A similar strategy was used in \cite{ZennaroAnguloPellejero2021} for the real space case.

\section{Validation}
\label{sec:validation}

In this section, we perform a series of tests to determinte the accuracy of our RSD clustering emulator. 

\subsection{A first look at the RSD emulator}

First, we examine the relationship between the emulated multipoles and the cosmological parameters. Fig.~\ref{fig:emuVSparams} displays the ratio of the monopole, quadrupole, and hexadecapole at a given cosmology with respect to a reference cosmology (defined by the middle points of the Eq.~\ref{eq:par_range} range set). Dashed lines show the same quantity but computed by \texttt{velocileptors}. On large scales, both our $N$-body-based emulator and the theory predictions coincide, indicating that the emulator captures the cosmology dependence accurately on those scales. When approaching smaller scales, $k\sim 0.1-0.2\ihMpc$ both predictions start to depart, which approximately coincides with the range of validity of perturbation theory. In contrast, our RSD emulator should accurately capture nonlinearities in clustering, even on scales $k \sim 0.6\ihMpc$. 
Regarding the cosmological parameter dependencies, $\Omega_{\rm m}$ and $\sigma_8$ exhibit an overall change in amplitude at all scales, demonstrating the dependence of the clustering amplitude on the amount of matter and the size of the perturbations. The parameters $\Omega_{\rm b}$ and $h$ have the greatest effect on the Baryon Acoustic Oscillations (BAO) scales, as one controls the baryon component and the other controls the units used to measure them (see e.g. \citealt{ShapeFit1}). The slope parameter $n_s$ displays the greatest dependence at large scales, while its effect is diminished at smaller scales by non-linearities. On the other hand, $w_0$ and $w_a$ show their biggest impact on small scales. Unfortunately, $m_{\nu}$ does not appear to have a significant effect on the RSD clustering in the tested range as compared to other parameters. This does not mean that we cannot measure this parameter; if the dependencies break some degeneracies in the parameter space, it is theoretically possible to detect their signal. We leave this for further work. Note that the quadrupole at the fiducial cosmology crosses zero at scales of  $k\approx0.6h/{\rm{Mpc}}$ (we did not include any FoG effect in this plot), which accentuates the differences we observe in this multipole at small scale ratios.

Interestingly, we see that small scales are sensitive to cosmologial parameters, most notably $\sigma_8$. Hence, we might expect stronger constraints on this parameter when including high $k$-modes. Indeed, we will later show this is the case.

\subsection{Error assessment}
\label{sec:errors}

Even though it is generally assumed that theoretical models have no associated errors because they are based on analytical computations that do not suffer from common sources of noise such as cosmic variance or shot noise, the reality is that theoretical predictions for a given observable are not exact but have inherent uncertainties. Specifically, for the case of models built from $N$-body simulations, uncertainties in the initial conditions (cosmic variance), arbitrariness in group finders, errors introduced by the finite accuracy of  force calculation and time integration, etc, will result in stochastic predictions. Similarly, in the case of perturbation theory, uncertainty can arise from the contribution of ignored orders, approximations in the equations of motions, and also from neglected physics such as galaxy formation (see eg. \citealt{Baldauf2016}). Our hybrid model will suffer from both sources of uncertainties. These ``theory'' errors can be taken into account by a suitable modification of the covariance matrix defining the likelihood of a given analysis \citep[see e.g.][]{Audren2013,Sprenger2019,Pellejero-Ibanez2020}. The main challenge is actually estimating this model covariance and determine whether it has a significant effect compared to the data covariance. 

To estimate the importance of the uncertainties in our RSD emulator, we first compare the scaling and neural network to the quantities they are simulating, and then we write the error budget against a well-known LSS survey error.

\subsubsection{Test Suite of Simulations}
\label{sec:suitesimulations}

The emulator presented in this work relies on employing the 4 BACCO simulations to sample hundreds of different cosmologies using a re-scaling algorithm (see \S\ref{sec:rescaling} for further details). Consequently, determining the precision of this technique for the quantities we will be simulating becomes crucial.   

With this goal, we will employ a suite of 36 paired $N$-body simulations. These simulations consists of $1536^3$ particles in a cubic volume of approximately $512\hMpc$ side (the simulations have different volumes to match those of the original simulations after rescaling). The volume explored by these simulations is smaller than our main BACCO simulations, but are defined to have identical numerical parameters, mass resolution, and force accuracy. These simulations were also carried out with the ``fixed \& paired'' methodology \citep{AnguloPontzen2016} with the latest version of {\tt L-Gadget3} and defining the initial conditions with second order Lagrangian Perturbation Theory (2LPT, see e.g. \citealt{BernardeauEtal2002}) at $z=49$. Moreover, these initial conditions have phases matching those of each of our BACCO simulations reducing significantly the role of cosmic variance.

The cosmologies chosen for our test suite vary one of eight cosmological parameters $\theta = \{\Omega_{\rm m}, \Omega_{\rm b}, \sigma_8, n_s, h, M_{\nu}, w_0, w_a \}$ covering the same range in which we build our emulator. When measuring the accuracy of our model predictions we compare the results of re-scaling this suite of simulations against those carried out directly with $N$-body codes at the target cosmology.

\begin{figure}
\includegraphics[width=\columnwidth]{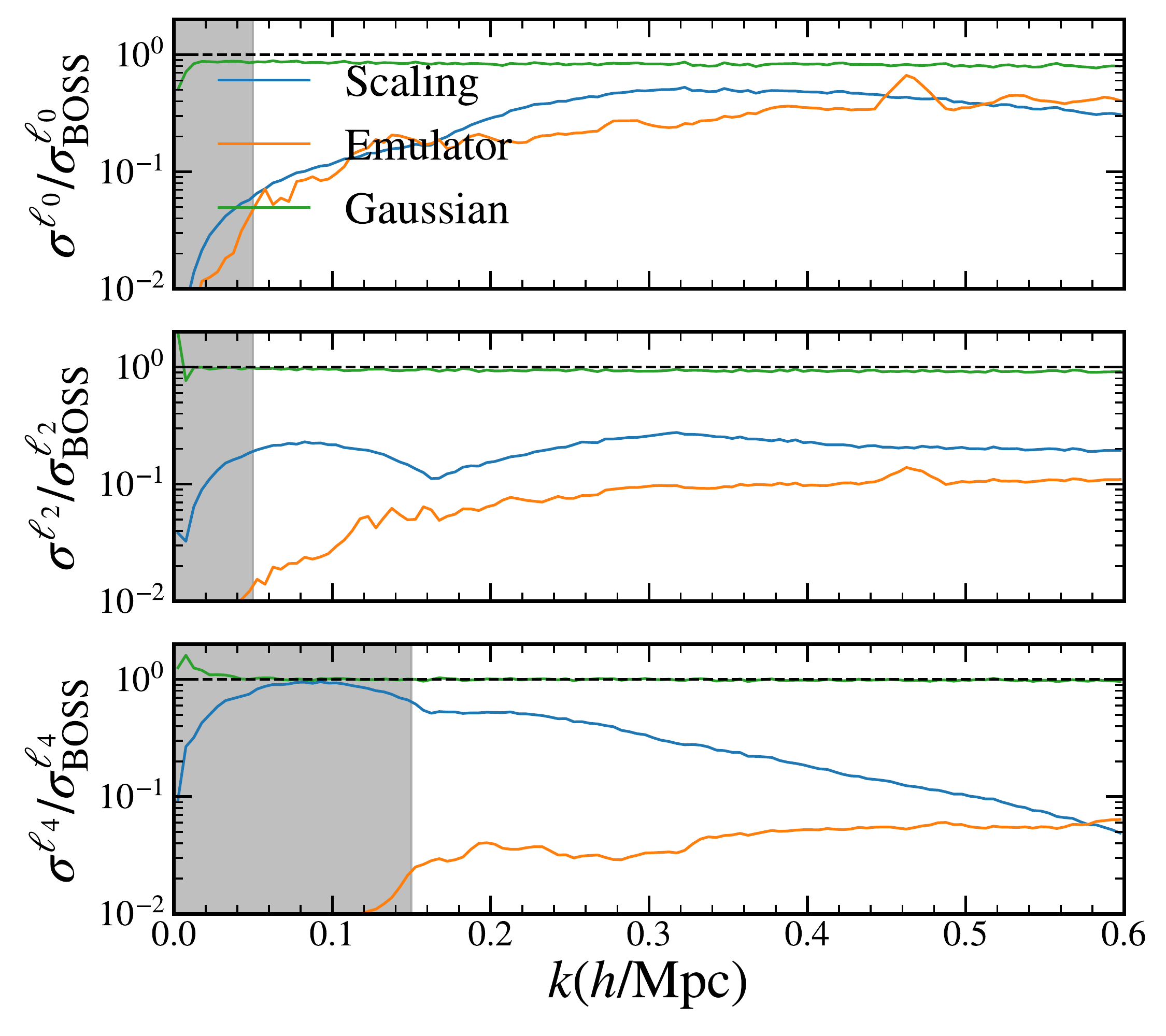}
\caption{Ratio of error elements (square root of diagonal covariance matrix elements) provided by the cosmology-scaling technique and the emulation with respect to the BOSS errors. Grey region represents the range of $k$'s in which we substitute our $N-$body emulator by the theory predictions. We further include the predictions Gaussian predictions for this number density and volume as green lines.  
\label{fig:error_budget}}
\end{figure}

\subsubsection{Uncertainties estimation}

We estimate the typical uncertainty in our emulated predictions for a given galaxy power spectrum, $P_\ell(k)$ as:

\begin{equation}
\begin{split}
  \sigma[P_\ell(k)] \sim \sigma \left[ \frac{P_{{\rm tr},\ell}^{\rm emu}-P_{{\rm tr,\ell}}^{\rm test}}{\sigma_{\rm G}[P_\ell^{\rm{test}}]} \right]  \sigma_{\rm G}[P_\ell] \, .
 \label{eq:diagerrors}
\end{split}
\end{equation}

\noindent where $P_{{\rm tr},\ell}$ are the multipoles of the power spectrum in the hybrid expansion (c.f. Eq.~\ref{eq:Pkmodel}), and the superscript ``emu'' and ``test'' refer to cases where we use cross spectra given by our emulator or computed directly in our our test suite of simulations, respectively. The first term is the standard deviation of this quantity computed over the test suite. When concerned with the error introduced by the cosmology scaling, this suite will correspond to the simulations presented in the previous subsection, whereas when concerned with emulation errors, the test suite will be a random subset of scaled simulations not employed in the training of the emulation. The $\sigma_G[P_\ell]$ expressions correspond to the Gaussian predictions for the cosmic variance uncertainty for $P_\ell$ \citep{Grieb:2015bia}, assuming the same volume as in our test suite. 

Essentially, this expression assumes that the typical uncertainty is independent of redshift and cosmology when expressed in units of the respective cosmic variance. In the case of real space, this is equivalent to assuming that the uncertainty is proportional to the power spectrum amplitude. However, in redshift space we cannot assume the uncertainty is proportional to the signal since the later crosses zero which makes the ratio ill defined. Although not shown here, we have indeed confirmed that the deviations between the emulated and test power spectra do not show clear cosmology nor redshift dependence.


\begin{figure*}
\includegraphics[width=\textwidth]{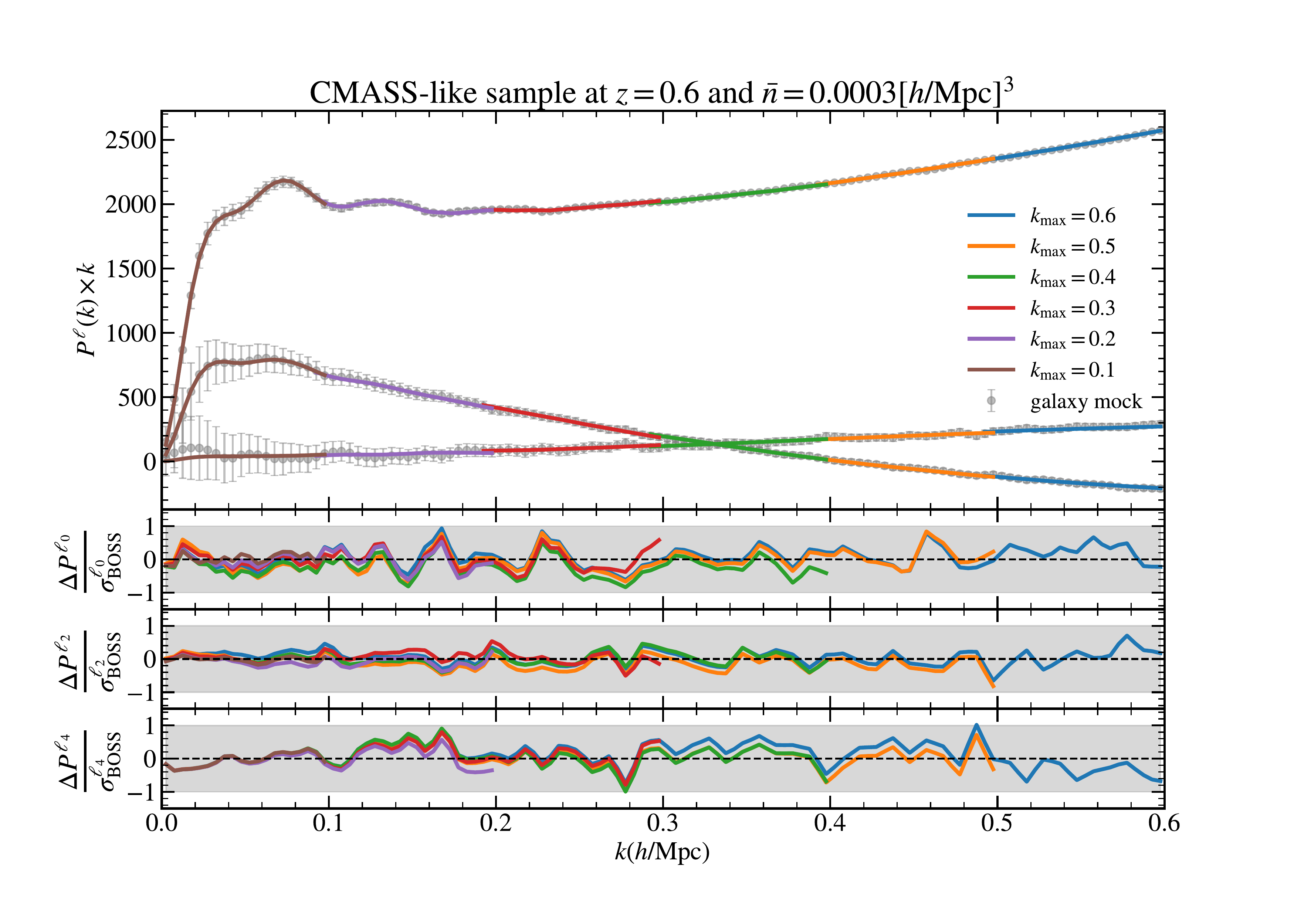}
\caption{Fit to the BOSS-CMASS galaxy mock. Grey dots represent the measured power spectrum from the galaxy mock with errorbars provided by the BOSS collaboration covariance matrix at $z\approx 0.6$. Different colours correspond to different $k_{\rm max}$ fits. Upper panel: monopole, quadrupole and hexadecapole measurements from galaxy mock and theory. Lower panels: fractional difference in units of the BOSS-CMASS diagonal errors. At all scales considered, the model can reproduce these two point clustering statistics. 
\label{fig:best_fit}}
\end{figure*}

In Fig.~\ref{fig:error_budget} we show our estimates for the uncertainty in our RSD emulator introduced by the cosmology rescaling and the neural network emulation as blue and orange lines, respectively. To compute such quantities, we have employed Eq.~\ref{eq:diagerrors} with a set of bias parameters that best describe the CMASS-BOSS sample at $z\sim0.6$ (see \S\ref{sec:best_fits}). We display these results in units of the diagonal elements of the covariance matrix of such sample, as computed by the BOSS collaboration\footnote{ \url{https://fbeutler.github.io/hub/deconv_paper.html}}. For comparison, we have also included as gren lines the Gaussian prediction for the CMASS sample.

We can see that the main source of uncertainty in our predictions, for all multipoles and scales, are those related to the cosmology-rescaling technique. However, this uncertainty is smaller than cosmic-variance uncertainties in the BOSS-CMASS sample. In the case of the monopole, it is at most 50\%, and typically much smaller. Note that, unfortunately, the error estimation is itself uncertain -- especially for the quadrupole and hexadecapole on large scales -- owing to the comparatively small volume of our suite of test simulations. Thus, our estimation is likely an overestimation of the true uncertainty. Nevertheless, these scales are given by theoretical predictions (indicated by shaded grey area), therefore, we can conclude that our RSD emulator is sufficiently accurate for the cosmological analysis of the CMASS survey. 

Finally, we note that the uncertainty assessment should be done for a given set of bias parameters (e.g. if a given sample has a $b_2\sim0$, the accuracy of $P_{2,2}$ becomes less relevant) and thus separately for a specific survey. This, combined with the larger cosmic volumes, implies that for the analysis of forthcoming surveys, it could be possible that an improved accuracy is necessary -- this could be easily achieved by including further simulations in the BACCO suite or by potentially improving the accuracy of the cosmology scaling technique.

\section{Application: Cosmology from a mock BOSS-CMASS sample}
\label{sec:application}

In this section, we use our emulator to extract cosmological constraints from a mock galaxy catalogue akin to BOSS-CMASS. Our goal is to assess the ability of our model to provide unbiased parameter estimates from a state-of-the-art observational survey.

\subsection{Stellar Mass selected galaxies}
\label{sec:galmock}

The SDSS-IV BOSS survey is currently the largest galaxy map of the Universe. BOSS delivered several galaxy catalogues, out of which CMASS provides the most accurate measurement of the large-scale structure. This corresponds to an effective average number density is $\bar{n}<4\times10^{-4}$[$h/$Mpc]$^{3}$ (see Fig.~2 of \citealt{BOSS2017}) and the selection roughly matches a stellar mass cut. Note that these numbers are effective values and that each redshift bin will have a lower number density. We will now describe our procedure to construct a mock galaxy catalogue that mimics CMASS.

We will create a catalogue in the TheOne-BOSS cosmology, defined in the last row of 
Table~\ref{tab:parameters_table}. This corresponds to the TheOne cosmology but with a smaller density fluctuations ($\sigma_8=0.8$). We do this so that the cosmological parameters are far from the boundaries of our emulator. To obtain this catalogue we scale the cosmology of the \texttt{TheOne} simulation. 

We then generate our galaxy mocks using the SHAMe (Sub Halo Abundance Matching extended) approach \citep{ContrerasAnguloZennaro2020AB, ContrerasAnguloZennaro2020}. SHAMe extends over the basic SHAM \citep[][]{ValeOstriker2006, ConroyWechslerKravtsov2006,Reddick2013,ChavesMontero2016,Lehman2017,Dragomir2018} by including orphans (subhaloes that are no longer resolved in a simulation, but that still could host galaxies), tidal disruption and a flexible amount of assembly bias. SHAMe accurately reproduces the real- and redshift-space clustering of stellar mass selected galaxies in the TNG300 magneto-hydrodynamic simulation (\citealt{10.1093/mnras/stx3040}, \citealt{10.1093/mnras/stx3304}, \citealt{10.1093/mnras/sty2206}, \citealt{10.1093/mnras/stx3112}, \citealt{10.1093/mnras/sty618}), thus it can be regarded as a realistic model for the spatial distribution of galaxies.

In SHAMe, the stellar mass of galaxies is given by five parameters, $\{\sigma_{M_*}, t_{\rm merger}, f_{\rm s}, A_{\rm c}, A_{\rm s}\}$, controlling the scatter in the considered abundance matching relation ($\sigma_{M_*}$), a variety of processes associated with the disruption of subhaloes and galaxies ($t_{\rm merger}, f_{\rm s}$), and the level of assembly bias of central and satellite galaxies ($A_{\rm{c}}$ and $A_{\rm{s}}$ respectively). We refer the reader to \cite{ContrerasAnguloZennaro2020} for a detailed explanation of these parameters and validation of the method, and \cite{ContrerasAnguloZennaro2020AB} for the assembly bias implementation.

To create a CMASS-like galaxy sample, we choose the SHAMe parameters $\{ 0.1786, 0.6617, 0.0073, 0.2404, -0.1163\}$ which have been tuned to fit the TNG300 simulation at a variety of number density cuts. We then select SHAMe galaxies with the largest stellar mass chosen to produce the number density $\bar{n}=3\times10^{-4}$[$h/$Mpc]$^{3}$ at redshift $z\approx 0.6$.

\subsection{Likelihood}

We fit the free parameters of our model using the Legendre multipoles of our mock CMASS galaxy power spectrum in redshift space. Specifically, we maximise the following likelihood:

\begin{multline}
 \log \LL = -\frac{1}{2} \left(P_\ell^{\rm gal}(k) - {P_\ell^{\rm emu}}(k)\right)^{\rm{T}}\,C_{\rm{BOSS}}^{-1}\,\left(P_\ell^{\rm gal}(k) - {P_\ell^{\rm emu}}(k)\right) \\ - \frac{1}{2}\log|C_{\rm{BOSS}}|- \frac{N}{2}\log(2\pi) \, , 
\label{equation:clust_like}
\end{multline}

\noindent where $N$ is the number of points in the data vector, $P_\ell^{\rm gal/emu}$ are the Legendre multipoles (c.f. Eq.~\ref{eq:pkpoles}) and $\ell=\{0,2,4\}$.

\begin{figure*}
\centering
\includegraphics[width=.45\textwidth]{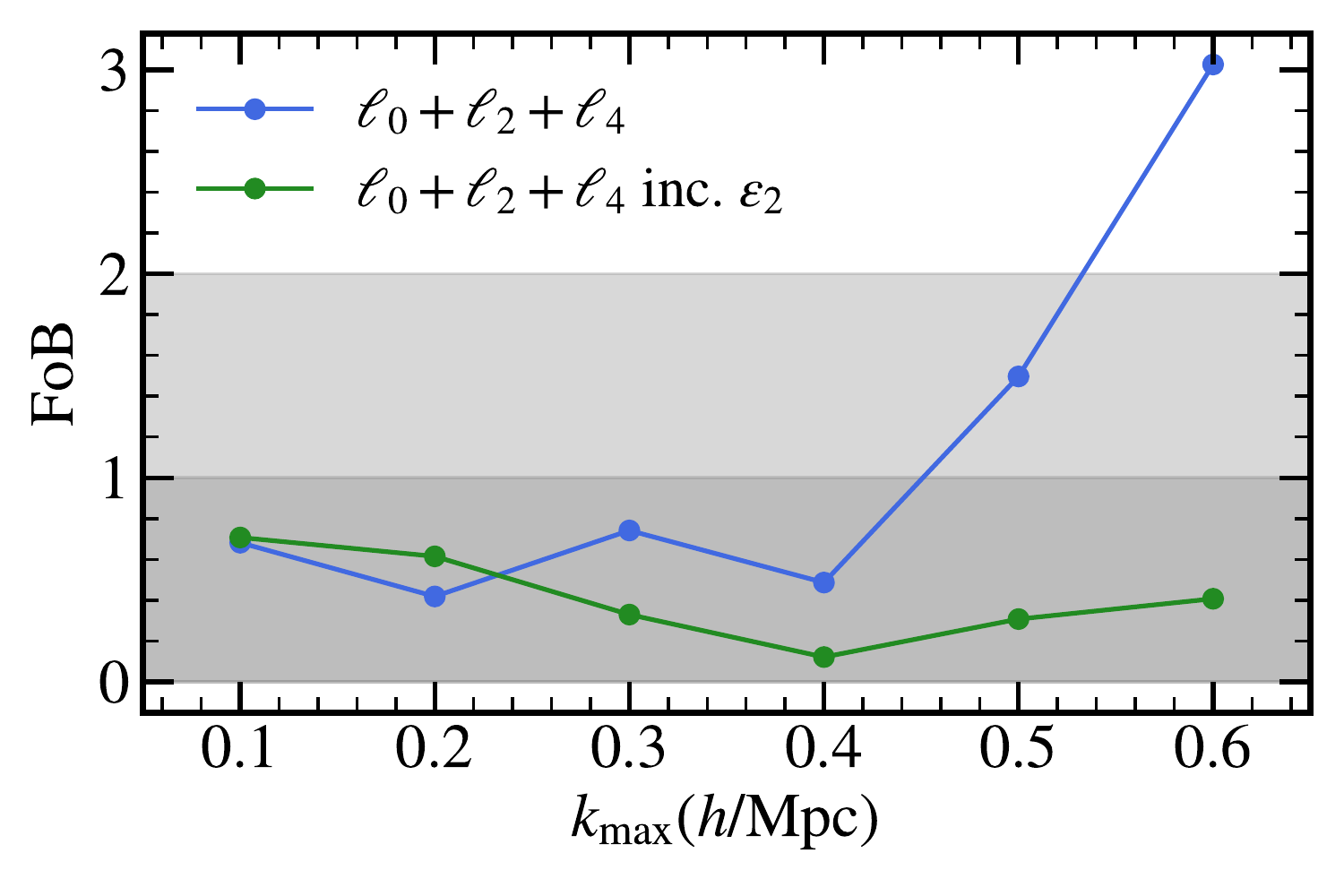}
\includegraphics[width=.45\textwidth]{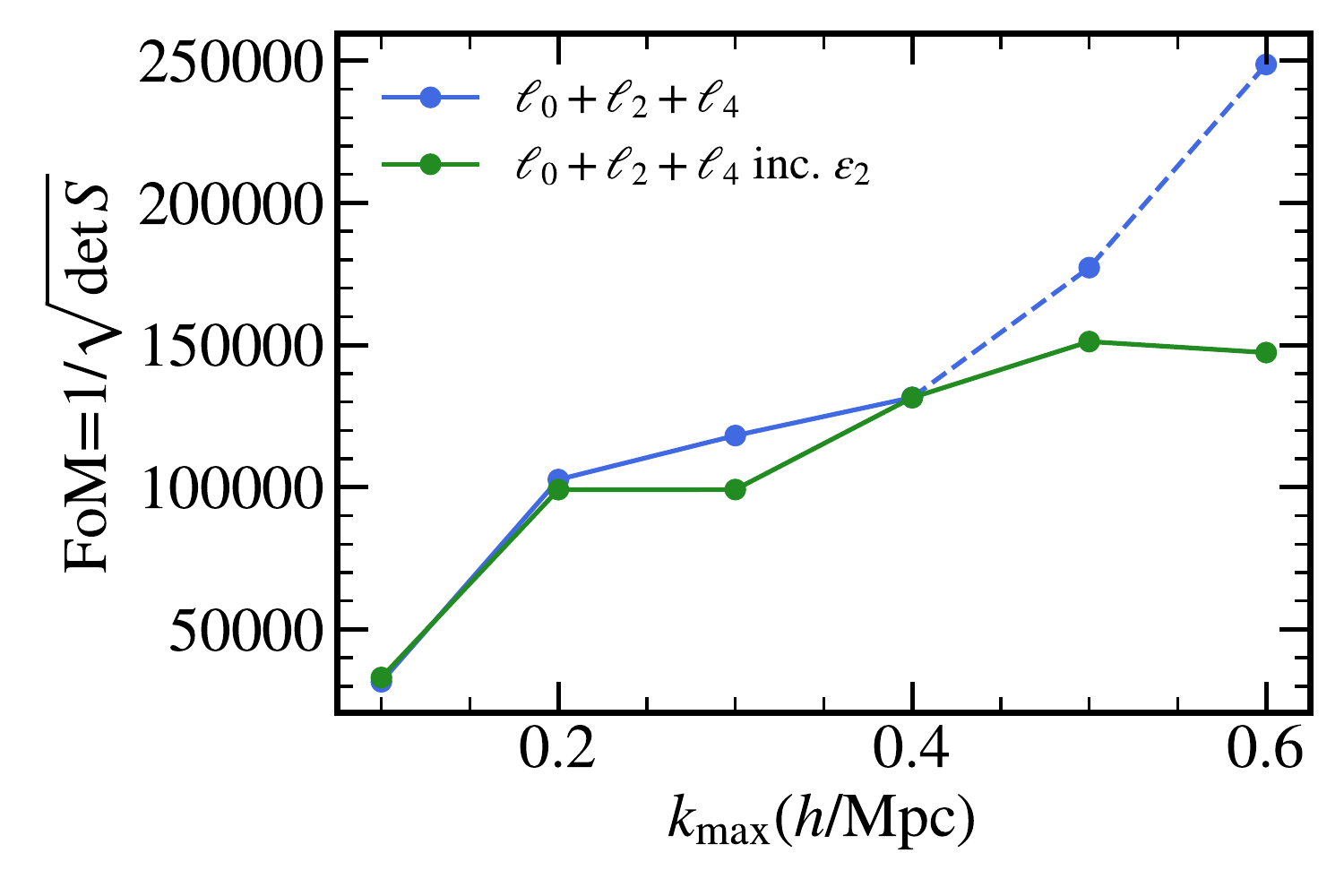}
\caption{Figure of Bias (left) and Figure of Merit (right) for different values of $k_{\rm max}$ for the set of cosmological parameters $\{\Omega_{\rm m},\sigma_8, h \}$. Green dots represent the model with a second-order Noise expansion (i.e. including $\epsilon_2$). Blue dots represent the model with a zeroth-order Noise expansion (i.e. with only $\epsilon_1$). This last model exhibits biases in relation to the actual parameters of the mock galaxy, resulting in priors effects that artificially increase the constraints. The dashed line in the FoM represents unreliable values contaminated priors.
\label{fig:FoB}}
\end{figure*}

We set $C_{\rm{BOSS}}[P_\ell(k),P_\ell(k')]$ as the covariance matrix estimated for the CMASS sample by the BOSS collaboration using PATCHY mocks \citep{Kitaura2016}. We note that these mocks have only been validated on scales $k>0.3\ihMpc$. Despite significant efforts -- see the works by the \texttt{BAM} team, \cite{Balaguera2019b, Balaguera2019a}, \cite{Pellejero2020a}, \cite{Sinigaglia2020}, and \cite{Kitaura2020} --, an accurate description of such a covariance matrix all the way down to $k\approx 0.6 [h/$Mpc] in redshift space has not yet been achieved. Nevertheless, we use the PATCHY estimate as a reasonable approximation to the true covariance. This is justified as we expect discreteness noise to be the dominant contribution for CMASS on scales $k>0.3\,[h/$Mpc] due to sparsity of the sample.

The diagonal elements of the BOSS-CMASS covariance are not very different from those of the Gaussian predictions (see e.g. \citealt{Lippich18,Colavincenzo18,Blot19}, and Fig.~\ref{fig:error_budget}), it is mainly in the non-diagonal terms -- dominated by the window function, wide angle effects and the nonlinearities -- where these covariances depart. In order to give an intuition of the magnitude of the uncertainties, in the range $[0.5h/{\rm{Mpc}}<k<0.6h/{\rm{Mpc}}]$, the errorbars of the BOSS-CMASS covariance account for about 0.4\% of the monopole, 10\% of the quadrupole, and 13\% of the hexadecapole's signal at a number density of 0.0003[$h/$Mpc]$^{3}$. We do not quote the results accuracy in these terms because when the signal approaches zero, these numbers diverge and lose interpretability.

To perform the fit, we fix the cosmological parameters to their true values except for $\{\Omega_m, \sigma_8, h\}$, leaving the full parameter space as $$\theta=\{\Omega_m, \sigma_8, h, b_1, b_2, b_{s^2}, b_{\nabla^2\delta}, \lambda_{\rm FoG}, f_{\rm sat},\epsilon_1, \epsilon_2\}.$$ In principle, we could have simultaneously varied all of the cosmological parameters listed in Eq.~\ref{eq:par_range}. Nonetheless, as shown in  Fig.~\ref{fig:emuVSparams}, we selected the three cosmological parameters with the greatest impact on $P_{\ell}$ so to avoid projection effects from unconstrained parameters. 

We note that, due to multiple reasons, we expect the differences between the model and the data to be much smaller than what is expected given $C_{\rm BOSS}$, especially on large scales. First, this covariance is affected by effects absent in our data vector: window function, wide angle effects, redshift evolution in clustering and selection function. Second, the Nyquist frequency used for the estimation of the 2-point statistics in the PATCHY mocks is $k_{\rm Nyq}\approx0.6[h/$Mpc], thus similar scales will be affected by aliasing. Finally, to reduce the variance as much as possible, the spectrum multipoles of the galaxy mock are computed from fixed amplitudes and paired $N$-body simulations, and averaged over 3 line-of-sight directions, $\{\hat{x}, \hat{y}, \hat{z}\}$. Therefore, we do not expect $\log{\mathcal{L}}$ to be distributed as a multivariate Gaussian with covariance $C_{\rm BOSS}$, but to display much smaller values. Nonetheless, this in fact will help us spot potential issues appearing in the model capabilities to reproduce the mock clustering data. 

We then either minimise Eq.~\ref{equation:clust_like} or compute contours through the \texttt{MULTINEST}\footnote{\url{https://github.com/farhanferoz/MultiNest}} Bayesian inference tool for recovering credibility intervals ( \citealt{multinest1}, \citealt{multinest2}, \citealt{multinest3}). With these fits, we will determine whether the emulator is capable of recovering the clustering of SM-selected galaxies within the accuracy of the BOSS-CMASS covariance with unbiased cosmological parameters.

\subsection{Best fits}
\label{sec:best_fits}

In Fig.~\ref{fig:best_fit} we display the mock CMASS multipoles as grey symbols and the best fit model as solid coloured lines for various values of the maximum wavemode included, $k_{\rm max}$. In the bottom panels we show the difference between data and model in units of the diagonal elements of the covariance matrix. 

At the level of precision demanded by the BOSS-CMASS covariance, our model is indistinguishable from the data down to scales of $k\approx 0.6h/{\rm{Mpc}}$ in the three multipoles. In particular, we see that the BAO feature is correctly captured. Additionally, as anticipated in the previous section, the mock data has much smaller scatter than that suggested by the error-bar. The superb agreement between data and model on small scales is a remarkable result by itself. However, this alone is not sufficient proof that an analysis with this model will give unbiased constraints of the cosmological parameters, for instance, an ingredient missing in the model could be compensated by changing the value of $\sigma_8$. We explore this next.

\subsection{Parameter inference}
\label{sec:param_inf}

\begin{figure}
\includegraphics[width=\columnwidth]{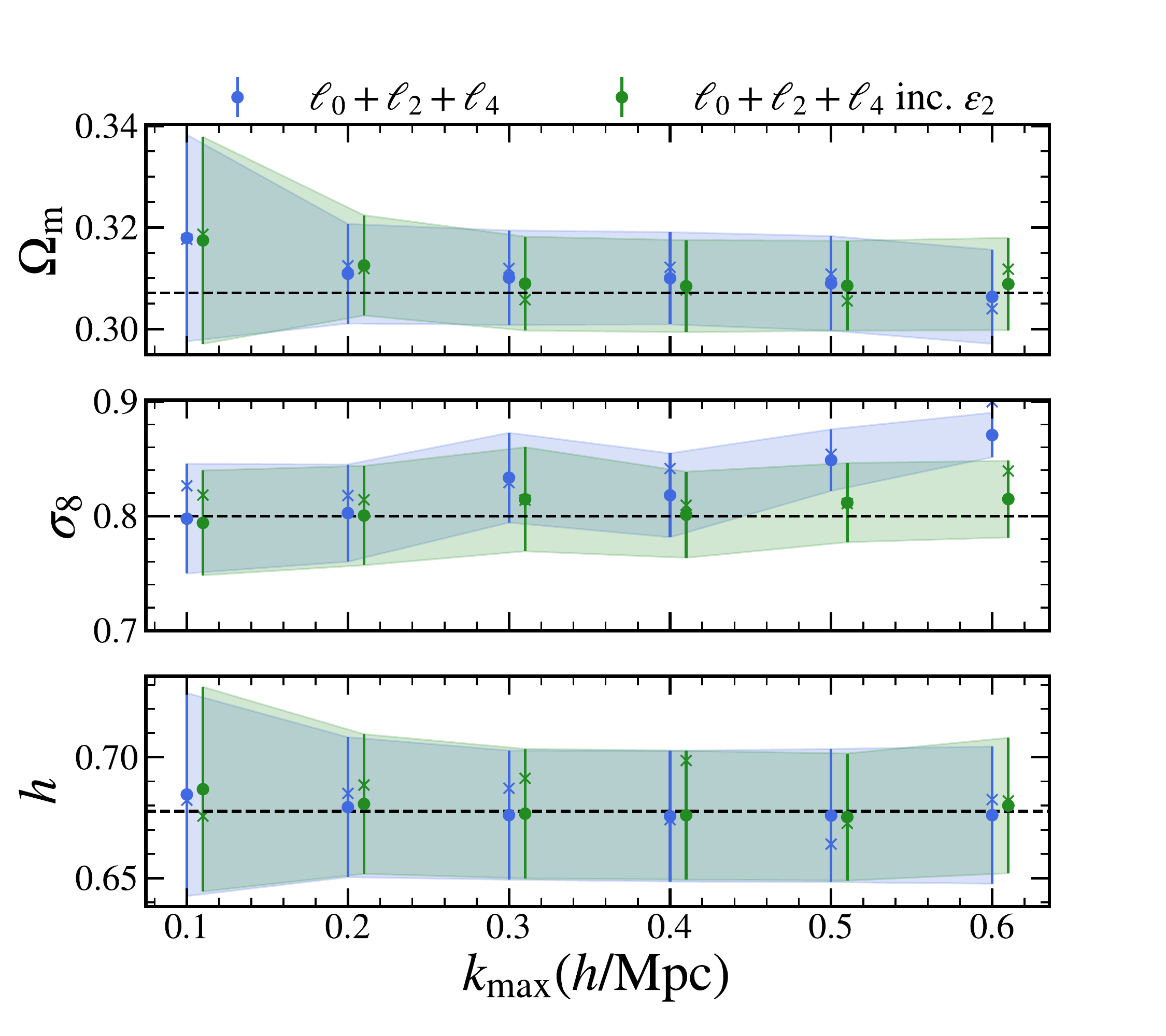}
\caption{ Cosmological parameter constraints derived from the Bayesian inference of comparing the model presented in this work with galaxy mock data under the assumption of a BOSS-CMASS covariance matrix. Dashed lines represent the actual galaxy mock values. The regions shaded in blue represent the results of a model that accounts for first order noise expansion (i.e. with only $\epsilon_1$). The regions shaded in green represent the outcomes of a model that accounts for second-order noise expansion, i.e. including $\epsilon_2$. Crosses indicate the best-fit maximum likelihood values. The green dots have been shifted slightly from their measured $k$ values to avoid clutter.  
\label{fig:cosmo_params}}
\end{figure}

In this subsection, we perform a full Bayesian analysis to obtain marginalised posterior probabilities on the free cosmological parameters of our model. We will start by quantifying the possible biases in parameters and the gain on constraining power, to then explore the posterior on our three cosmological parameters.

To quantify the performance of our model, we compute the Figure of Bias (FoB) which measures the distance in parameter space between the best-fit parameters, $\theta_{\rm{fit}}$ and their true values $\theta_{\rm{true}}$. Specifically, the FoB is defined as:

\begin{equation}
\rm{FoB}^2=\frac{(\theta_{\rm{fit}}-\theta_{\rm{true}})^{\intercal}S^{-1}(\theta_{\rm{fit}}-\theta_{\rm{true}})}{\chi^2_{p,1-\alpha}} \, .
\label{eq:FoB}
\end{equation}

\noindent where $\chi^2_{p,1-\alpha}$ is the $1-\alpha$ quantile of the $\chi^2_p$ distribution, and $S$ is the covariance matrix of parameters. Here we will focus on our three cosmological parameters, thus, we set $1-\alpha=32\%$ and $p=3$, and compute the denominator as the inverse of the cumulative distribution function of the $\chi^2$ distribution with 3 degrees of freedom. Therefore, FoB values that exceed unity can be interpreted as having a set of cosmological values in statistical tension with the true values.

We complement this measure with the Figure of Merit (FoM), which can be interpreted as the magnitude of the parameter information of a given setup. The FoM is defined as the inverse of the determinant of $S$:

\begin{equation}
\rm{FoM} = \frac{1}{\sqrt{\rm{det}\,S}} \, .
\label{eq:FoM}
\end{equation}

The values of the FoM and FoB as a function of the maximum scale included in our fits are shown in  Fig.~\ref{fig:FoB}. Blue lines show the results using a scale-independent noise term in our model whereas green lines include a higher order term, $\epsilon_2$ (c.f. Eq. \ref{eq:Noise}).

We can see that regardless of the noise modelling, our RSD emulator returns cosmological constraints that are unbiased at the 32\% level down to scales $k_{\rm max}=0.4\ihMpc$. On the right panel we show that employing those scales approximately have a FoM a factor of 4 larger than using a $k_{\rm max}=0.1\ihMpc$. For the constant noise model, including smaller scales leads to an important bias in our parameters -- the FoB reaches $\sim 3$. Due to those biases, out best fit values hit the boundaries of our emulator, thus the FoM is artificially increased (we highlight these FoM values are unreliable by displaying those results with dashed lines). In contrast, by employing an additional noise term, the FoB decreases essentially on all scales and it is well below unity even for $k_{\rm max}=0.6\ihMpc$. Interestingly, even though in those scales the shotnoise contribution is almost dominant, we appreciate a small but noticeable increase in the FoM of about 15\% compared to $k_{\rm max}=0.4\ihMpc$.

In Fig. \ref{fig:cosmo_params} we display the $1\sigma$ level of posterior probabilities of the three cosmological parameters after marginalising over all the free parameters of our model. Horizontal dashed lines mark the true values of the parameters. Consistent with the FoM results discussed above, we can see an important improvement in the the accuracy with which parameters are constrained when decreasing $k_{\rm max}$ from $0.1\ihMpc$ to $0.4\ihMpc$. When using small scales, we appreciate a strong bias in the recovered value of $\sigma_8$, however, when using a 2nd order stochastic contribution, the true underlying value is recovered. Despite the additional free parameter ($\epsilon_2$), constraints on $\Omega_m$ and $h$ are not degraded and there is still an improvement in $\sigma_8$ of about 25\% when using $k_{\rm max}=0.6\ihMpc$ compared to $k_{\rm max}=0.2\ihMpc$ -- the maximum wavenumber usually employed in these kind of analyses. 

It is interesting to explore whether the saturation of the cosmological information for $k_{\rm max}>0.4\ihMpc$ is due to the dominant role of shotnoise or due to an actual loss of information due to the nonlinear nature of small scales and the freedom imposed by galaxy formation (i.e.  small scales would constrain the value of bias parameters, not cosmological parameters). To investigate this we have repeated our analysis but employing a sample with 10 times the number density. We show these results in detail in Appendix~\ref{App:high_nmean}. In this case, our RSD emulator still delivers unbiased cosmological constraints (FoB<1) up to $k_{\rm max}=0.6\ihMpc$, but there is indeed a significant gain in cosmological information -- the FoM increases by roughly a factor of 2 from $k_{\rm max}=0.4$ to $0.6\ihMpc$ (to be compared to 15\% in our fiducial mock). As in our main CMASS mock, most of the cosmological information channels into improving the constraints on $\sigma_8$. This is somewhat expected in the light of the results shown by Fig.~\ref{fig:emuVSparams}, which indicated that the redshift-space multipoles on small scales are particularly sensitive to $\sigma_8$.

Another important validation is the scale independence of the best fit parameters. In Fig. \ref{fig:cosmo_params} we do not appreciate any systematic evolution of the posteriors with respect to $k_{\rm max}$. Additionally, in Appendix~\ref{App:scale_dep} we investigate the scale-dependence of the nuisance parameters of our model -- the bias, FoG, and stochastic terms. As for the cosmological parameters, we find that the recovered values are roughly scale independent as long as they are not affected by the boundaries of our emulator. In summary, the tests presented in this section indicate that our model is accurate enough for an analysis of realistic galaxy distributions.


\section{Summary and conclusions}
\label{sec:conclusion}

In this paper we present an emulator for the redshift-space galaxy power spectrum. The underlying model adopts a hybrid approach where the relation between galaxies and matter is given by a perturbative bias expansion whereas the matter statistics are given by the results of cosmological $N$-body simulations.

Our nonlinear matter statistics are obtained as a function of cosmology as follows. First, we compute the relevant statistics at 4000 combinations of cosmological parameters and redshifts using only 4 $N$-body simulations in combination with a cosmology-scaling technique. Second, we train a feed-forward neural network using these results to obtain new predictions at any point of our cosmological parameter space in approximately 0.3 seconds. Specifically, combined with a flexible model for small scale random velocities, we obtain predictions for the redshift space two-point statistics as a function of 8 different cosmological parameters, including massive neutrinos and dynamical dark energy. This RSD model can then be used in traditional Bayesian clustering analyses.

We quantify the uncertainty associated with our RSD emulator (for a fixed set of bias parameters that best fit a CMASS galaxy mock data) finding that it is approximately 20\% of the BOSS-CMASS observational uncertainty. The emulator errors are currently largest near the the boundaries of our parameter ranges, where the scaling is not as accurate and the neural network has fewer training points. We anticipate these errors will decrease in the future by employing a larger initial set of simulations, by potentially improving further the scaling error, and by using a denser sampling of cosmologies.

We test our emulator with a stellar mass-selected mock catalogue with number density $\bar{n}=3\times10^{-4}[h/$Mpc]$^{3}$ at $z=0.6$, mimicking the BOSS-CMASS observational sample. We fit the monopole, quadrupole, and hexadecapole of the redshift-space power spectrum, assuming the same covariance matrix as the CMASS sample. Our emulator is capable of fitting the mock data down to small scales with high precision while recovering unbiased cosmological constraints, even when using the full range of scales studied here ($k = 0.001-0.6\ihMpc$). We quantify the amount of information stored in scales by computing the Figure of Merit as a function of the maximum scale included in the fit. We find that the FoM for a CMASS-like sample increases by approximately 15\% when considering scales smaller than $0.4\ihMpc$, this translates to approximately 11\% stronger constraints on $\sigma_8$. Constraints on $\Omega_m$ and $h$ do not improve, which is consistent with the redshift-space multipoles on small scales being most sensitive to the $\sigma_8$ parameter. 

The amount of information that can be extracted from small scales is limited by the presence of a dominant shot noise contribution for a CMASS-like sample. After marginalisation over the nuisance parameters of our model (the bias, FoG, and noise terms) the monopole provides almost no additional information and all the gain arise from the quadrupole and hexadecapole. This situation is somewhat different for denser samples. In the case of a 10 times denser mock catalogue, there is considerable information and the FoM up to $k_{\rm max} \sim 0.6\ihMpc$ is approximately twice as large as that up to $0.4\ihMpc$, which translates into 40\% stronger constraints on $\sigma_8$

We conclude that our RSD emulator is a precise and accurate model for the multipoles of the redshift-space galaxy power spectrum. It provides an alternative venue for exploiting the measurements of current observed LSS datasets, such as the BOSS-CMASS sample, specially for extracting cosmological information from small scales in a robust manner. We will address this and other topics in forthcoming publications.


\section*{Acknowledgements}
The authors acknowledge the support of the ERC-StG number
716151 (BACCO). MPI acknowledges the support of the ``Juan de
la Cierva Formación'' fellowship (FJC2019-040814-I). SC acknowledges the support of the ``Juan de la Cierva Incorporaci\'on'' fellowship (IJC2020-045705-I). The authors also acknowledge the computer resources at MareNostrum and the technical support provided by Barcelona Supercomputing Center (RES-AECT-2019-2-0012 \& RES-AECT-2020-3-0014). We thank Rodrigo Voivodic for useful discussions. This is science, ``But this is a turtle''.

\section*{Data Availability}

 The data underlying this article will be shared on reasonable request to the corresponding author. The Neural Network emulator will be made public at \url{http://www.dipc.org/bacco} upon the publication of this article.


\appendix

\section{Results on a higher number density galaxy mock}
\label{App:high_nmean}

\begin{figure*}
\includegraphics[width=\textwidth]{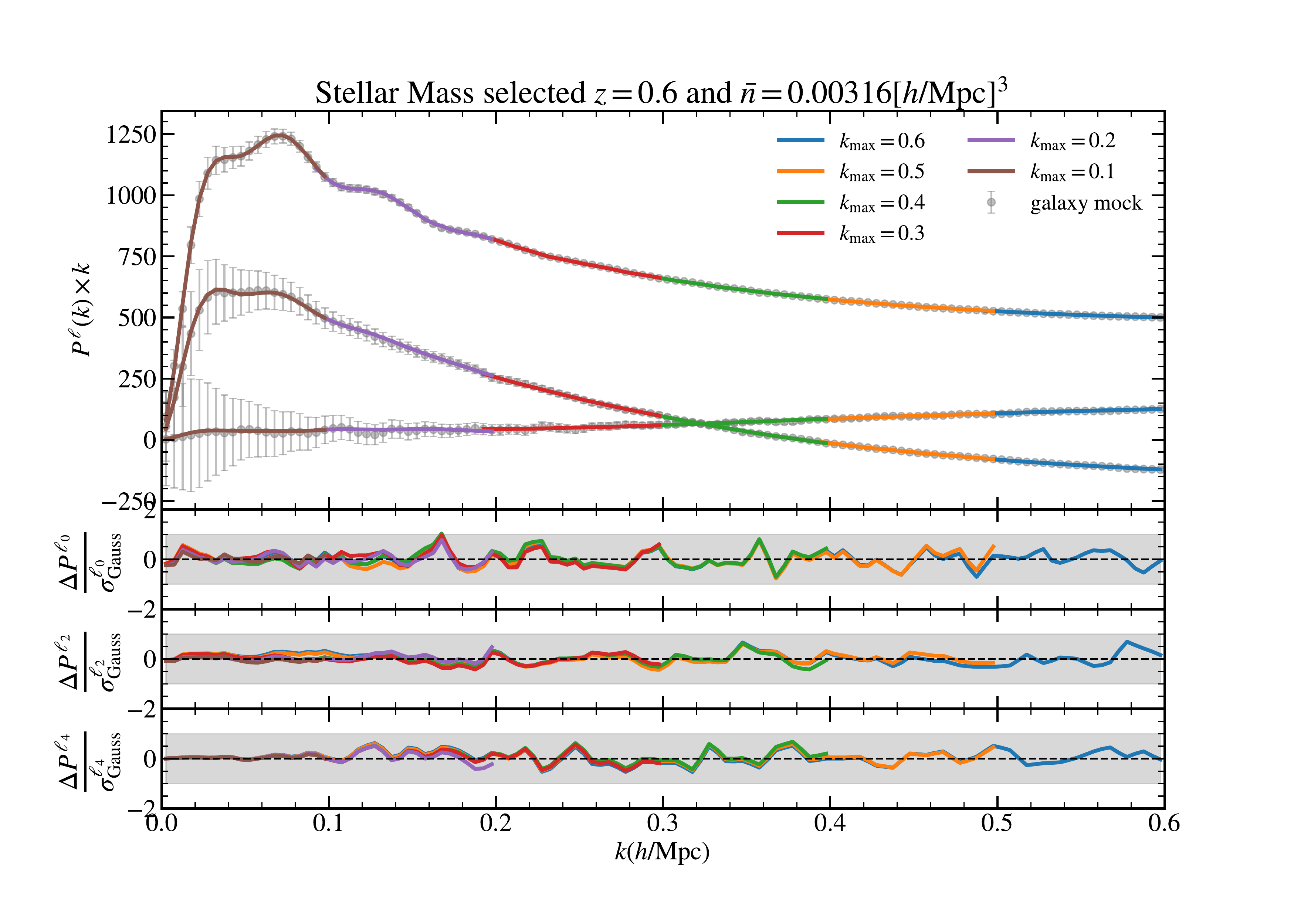}
\caption{Similar to Fig.~\ref{fig:best_fit} of the main text, except for a mock galaxy with 10 times the number density, $\bar{n}=0.00316$[$h/$Mpc]$^{3}$, and its corresponding Gaussian covariance matrix. Note that even at higher number densities with smaller errors the model still provides an accurate fit within the errors
}
\label{fig:best_fit_nmean}
\end{figure*}

\begin{figure}
\includegraphics[width=\columnwidth]{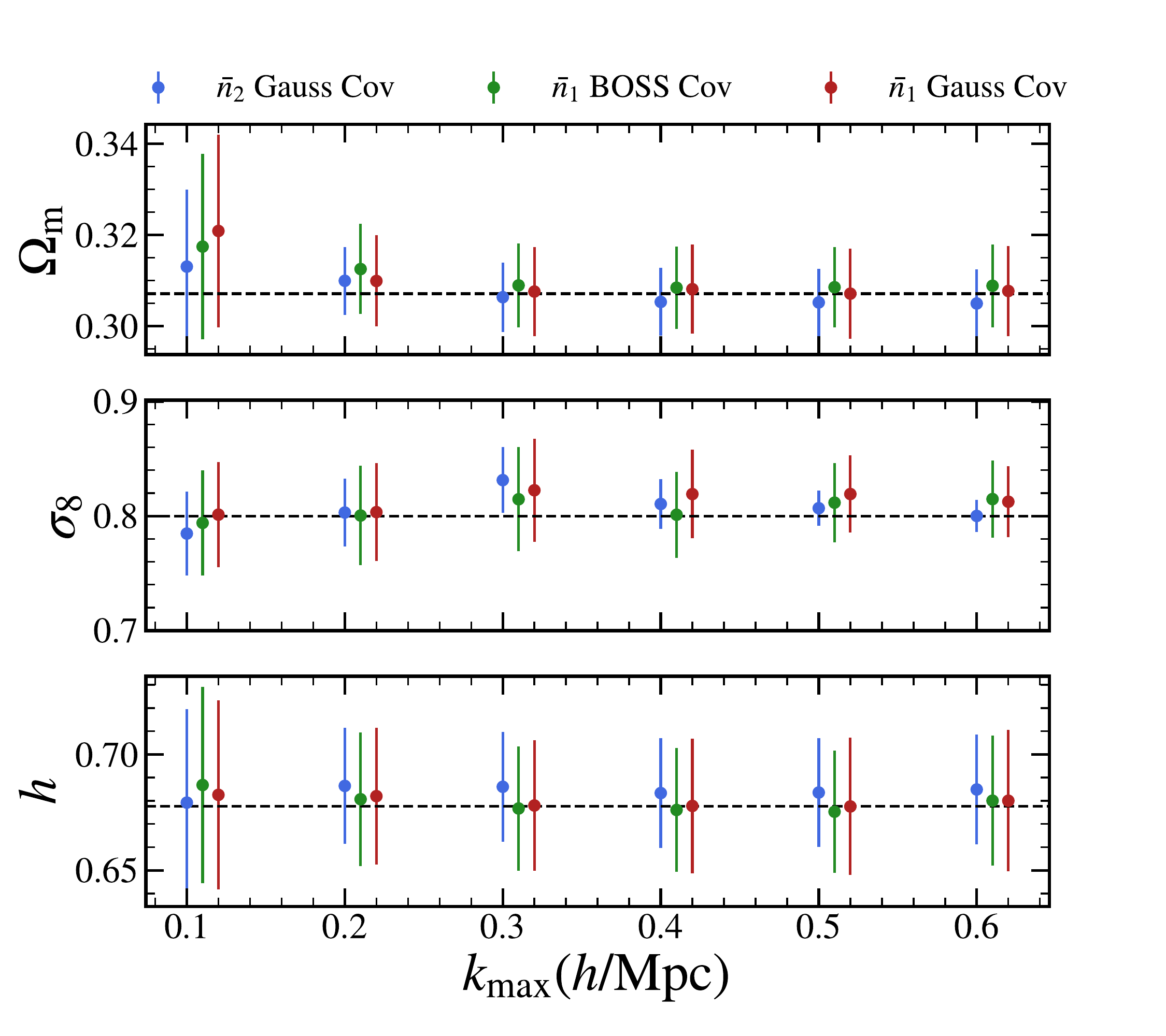}
\caption{ Same as Fig.~\ref{fig:cosmo_params} (green dots) of the main text ($\bar{n}_1=0.0003$[$h/$Mpc]$^{3}$) with the addition of the outcomes of fitting a galaxy mock with 10 times the number density ($\bar{n}_2=0.00316$[$h/$Mpc]$^{3}$) with a Gaussian covariance matrix (blue dots), and the outcomes of fitting the main text galaxy sample with the corresponding Gaussian covariance matrix (red dots) rather than the BOSS one.
\label{fig:cosmo_params_nmean}}
\end{figure}

In this appendix, we examine the performance of the emulator on a galaxy mock with ten times the number density of the galaxy examined in the main text. This examination has two objectives. The first objective is to evaluate the performance of the emulator on a completely different bias regime. Since the number density is greater, the number of tracers increases, thereby tracing the matter distribution more closely, and having a lower linear bias $b_1$. This causes the model to explore a different volume of the space of bias parameters. The second objective is to determine if the plateau observed on small scales in the right panel of Fig.~\ref{fig:FoB} can be explained by shot noise domination.

To build the galaxy mock we follow the procedure described in Sec.~\ref{sec:galmock}. We create the galaxy sample selecting the number of galaxies with the highest stellar mass content chosen to produce the number-density $\bar{n}=0.00316$[$h/$Mpc]$^{3}$ at redshift $z\approx 0.6$. We then compute the galaxy two-point multipoles of the ``Fixed \& Paired'' galaxy mocks over three orthogonal directions and take the average. We use these quantities to compute the associated Gaussian covariance matrix. Note that we cannot use the BOSS-CMASS covariance as in the main text because it was computed for a different number density regime. 

We show the results as grey dots with error-bars in Fig.~\ref{fig:best_fit_nmean}. The amplitudes of the multipoles are lower than in Fig.~\ref{fig:best_fit} due to the lower value of $b_1$. The smaller shot noise also pushes down the amplitude of the monopole at small scales. In this sample, the monopole signal is approximately 3 times that of the noise at the lowest $k$-bin. 

\begin{figure*}
\centering
\includegraphics[width=.45\textwidth]{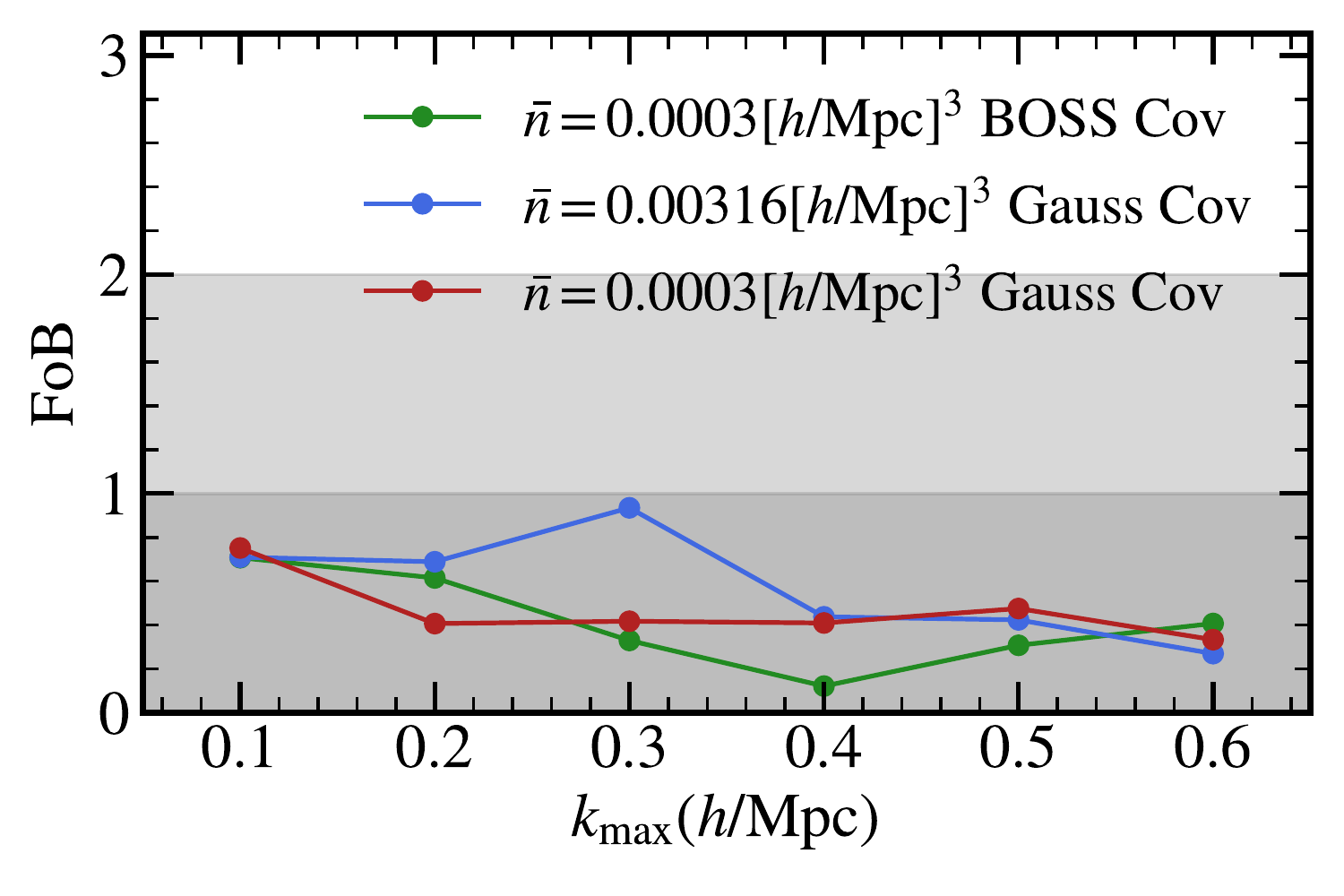}
\includegraphics[width=.45\textwidth]{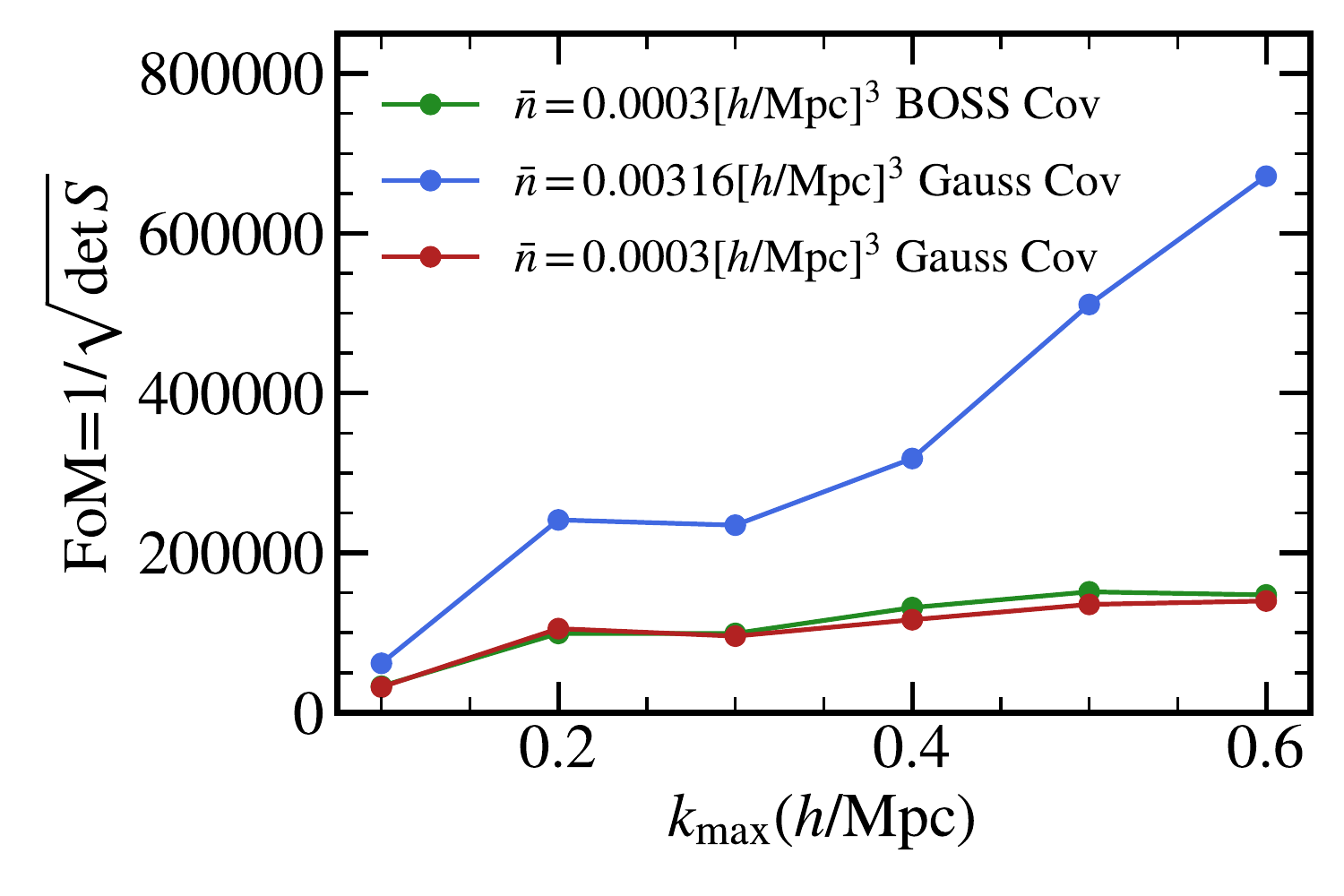}
\caption{Same as Fig.~\ref{fig:FoB} (green dots) of the main text ($\bar{n}=0.0003$[$h/$Mpc]$^{3}$ and BOSS covariance) with the addition of the results of a galaxy mock with 10 times the number density ($\bar{n}=0.00316$[$h/$Mpc]$^{3}$ with a Gaussian covariance matrix, blue dots), and the results of fitting the main text galaxy sample with its corresponding Gaussian covariance (red dots).  
\label{fig:FoB_nmean}}
\end{figure*}

We then fit the data for different values of $k_{\rm{max}}$ with our emulator (including $\epsilon_2$) and find the coloured lines of Fig.~\ref{fig:best_fit_nmean}. Once again, the model fits the data exceptionally well. There are no discernible deviations at any investigated scale. In addition, a complete Bayesian analysis reveals the projected statistics shown in Fig.~\ref{fig:cosmo_params_nmean}. We also identify unbiased constraints for this sample (blue dots). The FoB in the left panel of Fig.~\ref{fig:FoB_nmean} facilitates comprehension of this result. This FoB is the distance in the three-dimensional cosmological parameter space between the measured parameters and the actual parameters. The true parameters are always within the 1-$\sigma$ region of the measured ones, as found for the CMASS sample (green dots). This answers our initial question as to whether the emulator discovers the same results in a different volume of parameter space. A more thorough study with different cosmological parameters and galaxy mock parameters is left for future work.

From Fig.~\ref{fig:cosmo_params_nmean} we can also appreciate a significant reduction of the constraints with increasing values of $k_{\rm{max}}$, particularly in $\sigma_8$. The FoM depicted in the right panel of Fig.~\ref{fig:FoB_nmean} illustrates this result more clearly. When the scales from $k\approx 0.3h/{\rm{Mpc}}$ to $k\approx 0.4h/{\rm{Mpc}}$ are included, a plateau resembling that of the CMASS sample is observed. However, a substantial increase in constraining power is observed as scales become smaller. This points towards our initial hypothesis. In this regime, the shot noise does not dominate, and there is still a great deal of information to be gained. 

Nonetheless, we were compelled to use a Gaussian covariance for the high density sample because we lacked access to a more accurate estimation. The Gaussian approximation does not account for off-diagonal terms in the covariance coming from non-linearities, window function or wide-angle effects. These off-diagonal terms could potentially reduce the constraining power of the model and dominate over the shot noise effect. To have an intuition of their impact, we reproduce the study performed in the main text, but with the corresponding Gaussian covariance matrix. The results are shown in Fig.~\ref{fig:cosmo_params_nmean}, and Fig.~\ref{fig:FoB_nmean} as red dots. Non-diagonal terms appear to have little effect on the investigated cosmological parameters (even though we checked and the bias parameters are affected by this choice of covariance). We observe that $\sigma_8$ is better constrained when the covariance is Gaussian, but this effect is compensated by a smaller decrease in constraining power of the remaining cosmological parameters. 

We conclude, therefore, that the dominant difference on these small scales is shot noise. The shot noise erases the signal at scales greater than $k\approx 0.5h/{\rm{Mpc}}$ and as a consequence, there is no improvement in the constraints of the cosmological parameters.

\section{Scale dependence of bias parameters}
\label{App:scale_dep}

\begin{figure}
\includegraphics[width=\columnwidth]{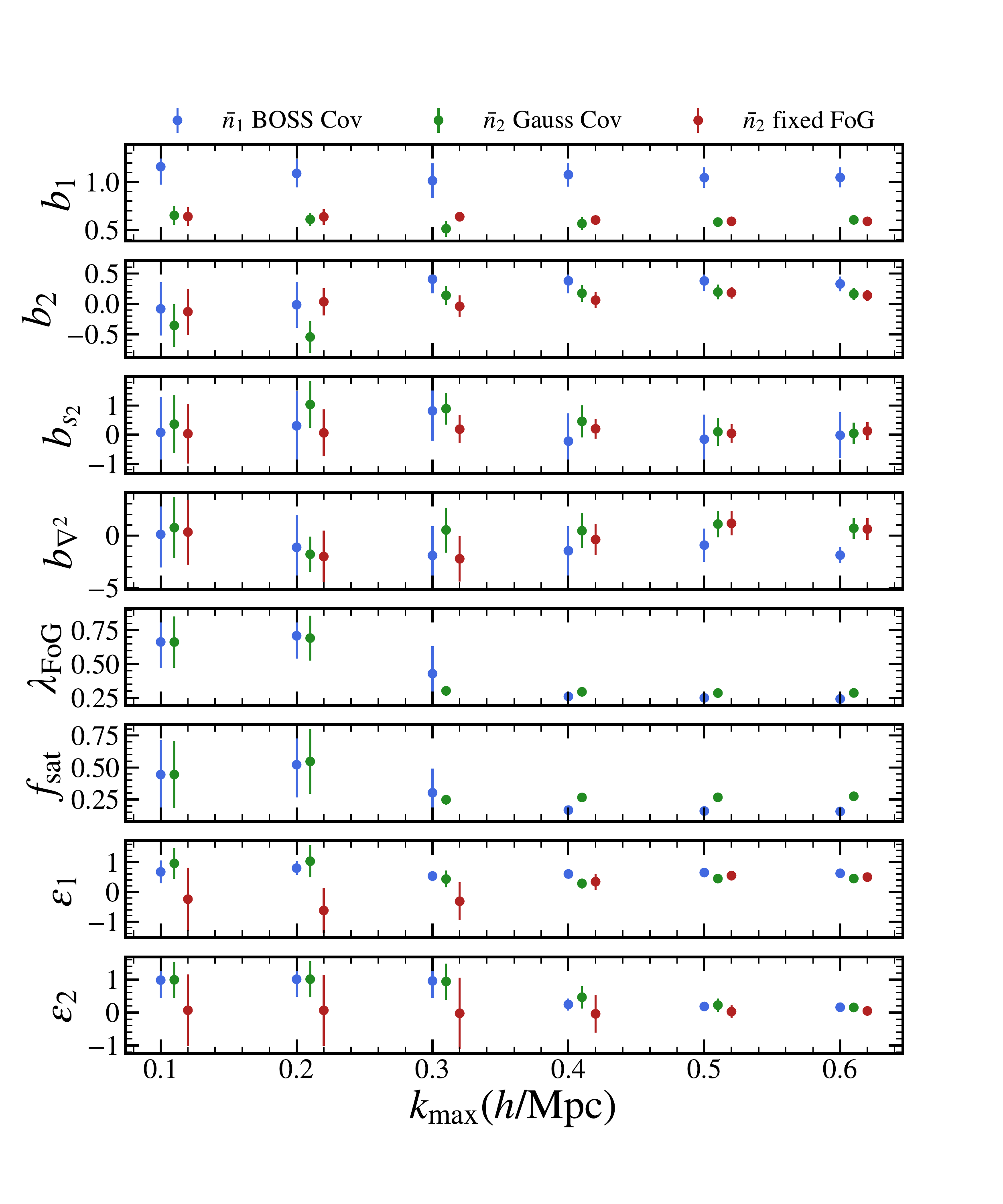}
\caption{ Bias parameters versus maximum scale of bayesian inference. Blue dots refer to the $\bar{n}_1=0.0003$[$h/$Mpc]$^{3}$ sample of the main text. Green dots represent the bias values of the $\bar{n}_2=0.00316$[$h/$Mpc]$^{3}$ sample in App.~\ref{App:high_nmean} studied with a Gaussian covariance. Red dots stand for the same $\bar{n}_2=0.00316$[$h/$Mpc]$^{3}$, except with fixed FoG parameters ($\lambda_{\rm{FoG}}$ and $f_{\rm{sat}}$) to the values found in the previous case at $k_{\rm max}= 0.6h/{\rm{Mpc}}$.
\label{fig:bias_params}}
\end{figure}

In this appendix we explore the behaviour of the bias parameters as a function of scale. We show in Fig.~\ref{fig:bias_params} the bias values found for the fit in Fig.~\ref{fig:best_fit} (blue dots). A priori, there is no clear tendency for the parameters shared between real and redshift space, $\{b_1,b_2,b_{s^2},b_{\nabla^2\delta},\epsilon_1, \epsilon_2\}$, to be scale-dependent. However, FoG parameters $\lambda_{\rm{FoG}}$ and $f_{\rm{sat}}$ exhibit a scale dependence for scales below $k_{\rm max}= 0.2h/{\rm{Mpc}}$ and above $k_{\rm max}= 0.3h/{\rm{Mpc}}$. This is due to how these parameters enter the model. According to Eq.~\ref{eq:densFoG}, the higher $\lambda_{\rm{FoG}}$, the lower its effect on the power spectrum. For values greater than 1, the Lorentzian function saturates, resulting in a complete degeneration above that threshold. Since $f_{\rm{sat}}$ enters as the amplitude of the exponential function, if the exponential becomes small (large values of $\lambda_{\rm{FoG}}$), $f_{\rm{sat}}$ will be unconstrained for large values and will avoid very small values that would cause the exponential term to become large again. Once $\lambda_{\rm{FoG}}$ identifies a $k$ scale above which it is constrained, the parameters become independent of scale (compare values above $k_{\rm max}= 0.3h/{\rm{Mpc}}$). This behaviour results in significant projection effects, which we observe as scale dependencies in the one-dimensional projections displayed here. We explore these projection effects in a clearer example with the following test.

Fig.~\ref{fig:bias_params} also displays the bias parameters found when fitting the higher number density sample of App.~\ref{App:high_nmean}. In this case, we observe a scale dependence in the $b_2$ parameter  (as well as in $\lambda_{\rm{FoG}}$) between scales below $k_{\rm max}= 0.2h/{\rm{Mpc}}$ and above $k_{\rm max}= 0.3h/{\rm{Mpc}}$. Note that there are three possible explanations: either the Gaussian covariance is a poor assumption for this sample and it is biasing our estimates, the projection effects on $\lambda_{\rm{FoG}}$ are affecting the $b_2$ estimate, or the model does not accurately describe this sample and absorbs dependencies not considered in $b_2$ from, for instance, higher orders in the bias expansion. We lack access to a more realistic covariance matrix at these scales, so we cannot rule out the first possibility completely. To determine whether projection effects influence our measurements, we fix the bias parameters describing FoG to the values found by fitting down to the smallest scales ($k_{\rm max}= 0.6h/{\rm{Mpc}}$) and conduct a similar fit. We find fits of comparable quality to those in Fig.~\ref{fig:best_fit_nmean}, as well as unbiased cosmological parameters. In Fig.~\ref{fig:bias_params}, the bias parameters identified by this test are depicted as red points. Note that the scale dependencies have vanished, indicating that when the ``true'' values of $\lambda_{\rm{FoG}}$ are assumed, the model is independent of scale. We therefore conclude that the primary source of the scale dependence observed in the case of higher number density is mainly due to projection effects arising from unconstrained FoG parameters. Therefore, there is yet no apparent need of additional orders in the bias expansion for the study of 2-point statistics in redshift space.  

Note that the bias expansion is being applied to smaller scales than it was originally intended for. Therefore, despite the fact that we used the same names for the parameters, there is no assurance that the values we obtain will coincide with those determined by perturbation theory techniques (see \citealt{ZennaroAnguloContreras2021} for extra details).

\bibliographystyle{mnras}
\bibliography{rsd} 







\bsp	
\label{lastpage}
\end{document}